\pdfoutput=1 
\PassOptionsToPackage{table}{xcolor}
\documentclass[5p,twocolumn]{elsarticle}

\usepackage{amssymb}
\usepackage{amsmath}
\usepackage{tcolorbox}
\usepackage{tabularx}
\usepackage{pifont}
\usepackage{multirow}
\usepackage{graphicx}
\usepackage{booktabs}
\usepackage{url}
\usepackage{balance}
\usepackage{float}

\AtBeginDocument{%
  }

\begin{document}

\title{Who Wins the Race? (R Vs Python) - An Exploratory Study on Energy Consumption of Machine Learning Algorithms}


\author[iitaddress]{Rajrupa Chattaraj}
\ead{cs22s504@iittp.ac.in}

\author[iitaddress]{Sridhar Chimalakonda}
\ead{ch@iittp.ac.in}

\author[accadress]{Vibhu Saujanya Sharma}
\ead{vibhu.sharma@accenture.com}
\author[accadress]{Vikrant Kaulgud}
\ead{vikrant.kaulgud@accenture.com}




\address[iitaddress]{Research in Intelligent Software \& Human Analytics (RISHA) Lab,\\
Department of Computer Science \& Engineering\\
Indian Institute of Technology Tirupati, India}
\address[accadress]{Accenture Labs, India}





\begin{abstract}
The utilization of Machine Learning (ML) in contemporary software systems is extensive and continually expanding. However, its usage is energy-intensive, contributing to increased carbon emissions and demanding significant resources. While numerous studies examine the performance and accuracy of ML, only a limited few focus on its environmental aspects, particularly energy consumption. In addition, despite emerging efforts to compare energy consumption across various programming languages for specific algorithms and tasks, there remains a gap specifically in comparing these languages for ML-based tasks. This paper aims to raise awareness of the energy costs associated with employing different programming languages for ML model training and inference. Through this empirical study, we measure and compare the energy consumption along with run-time performance of five regression and five classification tasks implemented in Python and R, the two most popular programming languages in this context. Our study results reveal a statistically significant difference in costs between the two languages in 95\% of the cases examined. Furthermore, our analysis demonstrates that the choice of programming language can influence energy efficiency significantly, up to 99.16\% during model training and up to 99.8\% during inferences, for a given ML task.
\end{abstract}






\maketitle

\section{Introduction}
\label{intro}

Artificial intelligence (AI) and Machine learning (ML) have become pervasive today, influencing various facets of our lives. Research indicates that over 77 percent of industrial companies have integrated AI into at least one of their operational functions\footnote{\url{https://explodingtopics.com/blog/companies-using-ai}}. The utilization of AI and ML technologies demands significant computational cost. According to a study done by MIT, the training of a single AI model can result in emissions exceeding 626,000 pounds of carbon dioxide which is  equivalent to almost five times the lifetime emissions of an average American car \cite{strubell2019energy}. This has prompted discussions concerning the environmental impacts and the imperative for energy-efficient solutions have stimulated research in the realm of Green Software Engineering (SE) \cite{manotas2016empirical}. Given the global reach and impact of AI and ML, it has become crucial for developers and practitioners to consider informed decision in energy consumption and the carbon footprint of software throughout its development and deployment processes.

Machine learning enthusiasts and data scientists commonly favor Python as their primary programming languages for extracting valuable insights from data. According to survey done in 2023, 69 percent of machine learning developers and data scientists now use Python \footnote{\url{https://adtmag.com/articles/2019/04/11/developer-economics.aspx\#}}. Its popularity in these fields can be credited to its developer-friendly syntax, the presence of comprehensive libraries like TensorFlow, PyTorch, and an active developer community \cite{nagpal2019python}. R also holds an important position as a popular programming language in the sphere of machine learning and statistics \cite{gupta2021performance}. Its strengths lie in the extensive range of packages it provides for tasks such as data manipulation and statistical analysis. While developers and practitioners often base their programming language choices on various features and characteristics \cite{georgiou2017analyzing}, it is also critical to consider the associated energy costs given their potential environmental impact.


Research has demonstrated that selection of programming languages \cite{georgiou2018your, couto2017towards} play a crucial role in determining the energy consumption of software components. Various programming languages exhibit differing levels of efficiency concerning speed, memory usage, and energy utilization \cite{pereira2017energy}. Studies have indicated that the choice of programming language can result in more energy-efficient algorithm implementations, while others may lead to less efficiency. For instance, opting for JavaScript over the Swift programming language has been found to potentially increase energy efficiency by up to 12,694\% \cite{georgiou2017analyzing}. While different programming languages have been compared for specific algorithms and tasks based on energy consumption, there has been a lack of evaluation concerning the energy consumption of programming languages within the context of machine learning algorithms and tasks.

This paper presents an in-depth empirical analysis to investigate and compare the energy consumption and run-time performance of various machine learning algorithms. The study includes five classification tasks and five regression tasks for both model training and inferences, utilizing the standard libraries of R and Python. In the context of ML, \textit{scikit-learn} stands out as the most popular Python library\cite{hao2019machine}, offering a comprehensive set of implementations for ML algorithms. \textit{CRAN} serves as a network of FTP and web servers globally, providing up-to-date versions of code and documentation for R\footnote{\url{https://cran.r-project.org}}, thus representing the R ecosystem broadly in the ML domain. \textit{scikit-learn} and CRAN are widely recognized as foundational resources for ML practitioners in Python and R,  respectively. Thus, we believe, comparing implementations from these two standard packages represent comparison of respective programming languages from an energy consumption perspective. However, as a further exploration, we also discuss our results on experimenting scratch-level implementations for Python. 






Our aim is to investigate \textit{which} programming language among Python and R, offers more energy and run-time efficient implementation for specific tasks during machine learning model training and inferences. Through this work, we aim to provide insights into the decision-making process regarding the selection of a programming language for specific machine learning tasks throughout their development and deployment stages. Our methodology consists of using various machine learning algorithms for classification and regression tasks, running them through model training and inference phases with 3 different datasets from UCI Machine Learning Repository\footnote{\url{https://archive.ics.uci.edu}} and Kaggle\footnote{\url{https://www.kaggle.com/datasets}}. Subsequently, we measure the energy consumption of each task independently. The energy consumption experiments were facilitated by utilizing energy consumption tools pyJoules\footnote{\url{https://pypi.org/project/pyJoules/}} (for Python) and RJoules \cite{chattaraj2023rjoules} (for R). Both of these tools are developed by leveraging the commonly used Intel-RAPL (Running Average Power Limit) interface \cite{khan2018rapl}. 


         \textit{To the best of our knowledge, this is the first empirical study that measures the  energy consumption and run-time performance of machine learning algorithms and tasks from a programming language perspective. We found that R is more energy and run-time efficient for regression algorithms, while Python is better for classification algorithms. Our results indicate that neither R nor Python is universally the most energy or run-time efficient choice across all the algorithms. Thus, ML engineers should carefully select the programming language based on specific tasks and contexts.}






 The rest of the paper is organized as follows. Section \ref{related_work} covers the Related Work. Research Questions are outlined in Section \ref{rq}, while Section \ref{tasks} specifies the addressed ML Tasks. The utilized Datasets are detailed in Section \ref{datasets}, and Section \ref{experiments} describes the Experimental Procedure. Findings from the experiments are provided in Section \ref{results}. Potential threats impacting the study's validity are discussed in Section \ref{threats}. Lastly, the Discussion and Conclusion are presented in Section \ref{discussion} and Section \ref{conclusion}, respectively.

 

\section{Related Work}
\label{related_work}

\def\checkmark{\tikz\fill[scale=0.4](0,.35) -- (.25,0) -- (1,.7) -- (.25,.15) -- cycle;}

In recent years, there has been a growing emphasis on examining the energy consumption of software components, leading to extensive exploration and analysis in diverse domains. The primary objective of these studies is to equip developers with valuable insights that aid in selecting energy-efficient options. A common theme in these research endeavors involves the assessment of energy consumption and the comparison of variations among the available choices. Various studies have demonstrated how different programming languages can influence the energy consumption of software\cite{pereira2017energy}, \cite{pereira2021ranking}, \cite{georgiou2018your}. Furthermore, factors like data structures\cite{calero2021investigating}, diverse frameworks\cite{oliveira2019recommending}, and dataframe libraries\cite{shanbhag2023exploratory} have shown their influence on energy consumption within software systems.



In the realm of machine learning, numerous researchers have dedicated their efforts to enhance energy efficiency. These methods encompass techniques such as model compression, including quantization \cite{lee2017lognet}, \cite{moons2016energy}, knowledge transfer through distillation \cite{sanh2019distilbert}, the utilization of efficient hardware accelerators \cite{chen2016eyeriss}, \cite{park2018energy}, and the implementation of energy-efficient computation methods in hardware \cite{jiao2018energy}, \cite{sarwar2018energy}. In this context, Schwartz et al. \cite{schwartz2020green} introduced a division between GreenAI and RedAI, promoting an emphasis on research dedicated to enhancing energy efficiency, rather than solely focusing intensely on surpassing state-of-the-art accuracy metrics. Recently, Duran et al. has introduced \textit{GAISSALabel}, a web-based tool designed to evaluate and label the energy efficiency of ML models. Additionally, GreenLight \cite{rajput2024greenlight} is a tool that provides energy consumption data along with relevant factors such as input parameter size, offering deeper insights into the energy usage of machine learning models. Despite the available techniques and studies investigating energy consumption in this domain, the impact of programming languages on energy usage during ML tasks remains largely unexplored.


Table \ref{tab:my-table} presents a summary of the relevant literature reviewed in the context of this study. Existing research predominantly concentrates on energy consumption, programming languages, or machine learning tasks individually or in paired combinations, as outlined in Table \ref{tab:my-table}. In \textbf{contrast}, our study delves into examining the relationship between \textbf{energy consumption} and run-time performance with \textbf{programming languages across various machine learning tasks}.  We assessed the energy consumption and run-time of 5 regression and 5 classification tasks using \textit{scikit-learn} implementations from Python and \textit{CRAN} repository from R language, for model training and inference phases. With this study we aim  to underscore the importance within the research community of investigating the energy efficiency of different programming languages within the machine learning pipeline.

\begin{table*}[]
\small
\centering
\caption{Summary of Existing Literature on Energy Consumption (EC) in Programming Languages (PL) and Machine Learning (ML) Tasks}
\label{tab:my-table}
\resizebox{\textwidth}{!}{
\begin{tabular}{|p{3cm}|p{9cm}|p{1cm}|p{1cm}|p{1.5cm}|} 
\hline
\rowcolor[HTML]{EFEFEF} 
\textbf{Studies} & \textbf{Focused Area} & \textbf{EC} & \textbf{PL} & \textbf{ML Tasks} \\ \hline

Hasan et al. \cite{hasan2016energy} & Analyzed energy profiles of various Java collection classes, revealing that opting for an inefficient collection could result in up to a 300\% increase in energy consumption. & \ding{51} &  &  \\ \hline

Weber et al. \cite{weber2023twins} & Analyzed the correlation between energy consumption and runtime performance among 14 real-world software systems. & \ding{51} &  &  \\ \hline

Schuler et al. \cite{schuler2020characterizing} & Explored the relationship between System API utilization and energy consumption in third-party software libraries. & \ding{51} &  &  \\ \hline

Shanbhag et al. \cite{shanbhag2023exploratory} & Studied the energy consumption of various dataframe processing libraries for data-oriented stages in ML pipeline. & \ding{51} &  & \ding{51} \\ \hline

Ournani et al. \cite{ournani2021comparing} & Compared the energy consumption of 27 Java I/O methods across different file sizes. & \ding{51} &  &  \\ \hline

Pereira et al. \cite{pereira2017energy} & Conducted a comprehensive analysis of energy efficiency across 27 different programming languages and investigated how energy consumption correlates with speed and memory utilization. & \ding{51} & \ding{51} &  \\ \hline

Georgiou et al. \cite{georgiou2017analyzing} & Conducted an empirical analysis on the energy consumption of eight different programming languages using the Rosetta Code Repository, demonstrating variations in energy efficiency. & \ding{51} & \ding{51} &  \\ \hline

Kumar et al. \cite{kumar2019energy} & Scrutinized the energy consumption of Java command line options, concluding that Oracle JDK demonstrated higher energy efficiency compared to Open JDK. & \ding{51} &  &  \\ \hline

Abdulsalam et al. \cite{abdulsalam2014program} & Evaluated the energy effect of memory allocation choices and the result indicated that malloc is the most efficient in terms of energy and performance. & \ding{51} &  &  \\ \hline

Maleki et al. \cite{maleki2017understanding} & Impact of object-oriented programming (OOPs) design patterns and the use of overloading and decorator on energy efficiency. & \ding{51} &  &  \\ \hline

Verdecchia et al. \cite{verdecchia2022data} & Carried out an empirical investigation regarding the energy consumption of various machine learning algorithms. & \ding{51} &  & \ding{51} \\ \hline

Georgiou et al. \cite{georgiou2022green} & Probed the energy consumption of deep learning frameworks such as TensorFlow and PyTorch, during the training and inference phases. & \ding{51} &  & \ding{51} \\ \hline

Lima et al. \cite{lima2016haskell} & Analyzed the energy efficiency of Haskell programming language and found that choosing right data sharing primitive can lead to 60\% more energy efficient solutions. & \ding{51} & \ding{51} &  \\ \hline

Xu et al. \cite{xu2023energy} & Studied the relationship between deep learning model architectures and energy consumption during training of neural networks. & \ding{51} &  & \ding{51} \\ \hline

Procaccianti et al. \cite{procaccianti2016empirical} & Empirically analyzes the energy consumption of software applications before and after applying two Green Software practices—query optimization in MySQL Server and usage of "sleep" instruction in the Apache web server—to evaluate their effectiveness in achieving energy savings and analyze resource-level trade-offs. & \ding{51} &  & \\ \hline

Nahrstedt et al. \cite{nahrstedt2024empirical} & Empirically examined the Pandas and Polars libraries for data operations and recommend using Polars for energy-efficient data analysis, highlighting the importance of CPU utilization in selecting libraries. & \ding{51} &  & \ding{51} \\ \hline
Khomh et al. \cite{khomh2018understanding} & Compares the individual and combined impact of six cloud design patterns on the energy consumption of multi-processing and multi-threaded applications deployed in the cloud environment, while also examining the trade-offs between application performance and energy efficiency. & \ding{51} &  & \\ \hline
\end{tabular}}
\end{table*}

\vspace{-3mm}
\section{Research Questions}
\label{rq}

The machine learning (ML) life cycle involves multiple stages, primarily model training and inferences, also known as the development and deployment phases \cite{ghoroghi2022advances}. During model training, an ML algorithm learns from training data, while inferences involve applying the algorithm to live data to make predictions \cite{georgiou2022green}. In the field of ML, two primary prediction problems often encountered are classification and regression \cite{kirchner2015classification}. While much research has focused on improving model accuracy and performance across these tasks, the energy consumption associated with different stages of the ML life cycle—particularly with respect to the underlying programming environment—remains underexplored. To address this gap, our aim is to investigate the impact of Python and R programming languages on energy consumption during model training and inferences. In order to achieve this, we formulate the following research questions (RQs).

\begin{enumerate}

\item \textbf{RQ1:} \textit{Which programming language, Python or R, is more energy and run-time efficient for regression tasks?}

\begin{itemize}
     \item \textbf{RQ1.1:} \textit{Which language is more energy and run-time efficient during the training phase for regression tasks?}
    \item \textbf{RQ1.2:} \textit{Which language is more energy and run-time efficient during the inference phase for regression tasks?}
\end{itemize}


\item \textbf{RQ2: } \textit{Which programming language, Python or R, is more energy and run-time efficient for classification tasks?}
\begin{itemize}
    \item \textbf{RQ2.1: }  \textit{Which language is more energy and run-time efficient during the training phase for classification tasks?}
    \item \textbf{RQ2.2: } \textit{Which language is more energy and run-time efficient during the inference phase for classification tasks?}

\end{itemize}



\end{enumerate}

\section{Machine Learning Tasks}
\label{tasks}

In the field of machine learning (ML), two primary prediction problems that are often encountered are classification and regression\cite{kirchner2015classification}. This study encompasses a diverse and popular\cite{das2017survey, luo2020comparing} set of machine learning algorithms, covering five regression and five classification tasks as described below.


\begin{table*}[]
\small
\centering
\caption{Details of Libraries Chosen for Machine Learning Tasks}
\vspace{2mm}
\resizebox{\textwidth}{!}{%
\begin{tabular}{|l|l|ll|ll|}
\hline
\multirow{2}{*}{}                     & \multirow{2}{*}{\textbf{ML Tasks}} & \multicolumn{2}{l|}{\textbf{Python( version=3.8.10)}}                                                                                & \multicolumn{2}{l|}{\textbf{R(version=3.6.3)}}                                                             \\ \cline{3-6} 
                                      &                                    & \multicolumn{1}{l|}{\textbf{\begin{tabular}[c]{@{}l@{}}sklearn Packages\\ (version=1.2.0)\end{tabular}}} & \textbf{Model}           & \multicolumn{1}{l|}{\textbf{\begin{tabular}[c]{@{}l@{}}CRAN\\ Libraries\end{tabular}}} & \textbf{versions} \\ \hline
\multirow{5}{*}{Regression Tasks}     & Linear Regression                  & \multicolumn{1}{l|}{linear\_model}                                                                        & LinearRegression         & \multicolumn{1}{l|}{glmnet}                                                            & 4.1.8             \\ \cline{2-6} 
                                      & Gaussian Regression                & \multicolumn{1}{l|}{gaussian\_process}                                                                    & GaussianProcessRegressor & \multicolumn{1}{l|}{MASS}                                                              & 7.3.51.5          \\ \cline{2-6} 
                                      & Decision Tree                      & \multicolumn{1}{l|}{tree}                                                                                 & DecisionTreeRegressor    & \multicolumn{1}{l|}{tree}                                                              & 1.0.43            \\ \cline{2-6} 
                                      & Support Vector Machine             & \multicolumn{1}{l|}{svm}                                                                                  & SVR                      & \multicolumn{1}{l|}{e1071}                                                             & 1.7.13            \\ \cline{2-6} 
                                      & Neural Network                     & \multicolumn{1}{l|}{neural\_network}                                                                      & MLPRegressor             & \multicolumn{1}{l|}{neuralnet}                                                         & 1.44.2            \\ \hline
\multirow{5}{*}{Classification Tasks} & Logistic Regression                & \multicolumn{1}{l|}{linear\_model}                                                                        & LogisticRegression       & \multicolumn{1}{l|}{glmnet}                                                            & 4.1.8             \\ \cline{2-6} 
                                      & Gaussian Naive Bayes                        & \multicolumn{1}{l|}{naive\_bayes}                                                                         & GaussianNB               & \multicolumn{1}{l|}{naivebayes}                                                        & 0.9.7             \\ \cline{2-6} 
                                      & Decision Tree                      & \multicolumn{1}{l|}{tree}                                                                                 & DecisionTreeClassifier   & \multicolumn{1}{l|}{rpart}                                                             & 4.1.21            \\ \cline{2-6} 
                                      & Support Vector Machine             & \multicolumn{1}{l|}{svm}                                                                                  & SVC                      & \multicolumn{1}{l|}{e1071}                                                             & 1.7.13            \\ \cline{2-6} 
                                      & Random Forest                      & \multicolumn{1}{l|}{ensemble}                                                                        & RandomForestClassifier   & \multicolumn{1}{l|}{randomForest}                                                      & 4.6.12            \\ \hline
\end{tabular}%
}

\label{tab:my-table1}
\end{table*}

\begin{table*}[h]
\caption{Dataset Information}
\vspace{2mm}
\label{tab:datasets-info}
\resizebox{\textwidth}{!}{%
\begin{tabular}{lllll}
\toprule
                    \textbf{Dataset Name} &                       \textbf{Source} & \textbf{No. of Data Points} &          \textbf{Tasks} & \textbf{No. of Attributes} \\
\midrule
                   Adult Dataset &  UCI Machine Learning Repository &             48,842 & Classification, Regression &                14 \\
             Drug Review Dataset &  UCI Machine Learning Repository &            215,063 & Classification, Regression &                 6 \\
New York City Taxi Trip Duration &                          Kaggle &          1,458,644 &     Classification, Regression &               11 \\
\bottomrule
\end{tabular}%
}
\end{table*}

\begin{itemize}

\item \textbf{Regression Tasks: } Regression tasks in ML involve predicting continuous numerical values based on input data. These tasks also aid in understanding the relationships between variables and making predictions based on historical data \cite{maulud2020review}. In this study, we have considered 5 widely used\cite{rodriguez2015machine, sharifzadeh2019machine} ML algorithms for regression tasks namely Linear Regression, Gaussian Regression, Neural Network, Decision Tree, and Support Vector Machine. 

\item \textbf{Classification Tasks: }Classification involves the task of identifying a model or function that aids in the division of data into distinct categorical classes, which are essentially discrete values \cite{kotsiantis2007supervised}. In this context, data is assigned specific labels based on input parameters, and the objective is to predict these labels for the given data. For classification tasks, our analysis incorporates popular algorithms\cite{nhu2020shallow} namely Logistic Regression, Gaussian Naive Bayes, Decision Tree, Support Vector Machine, and Random Forest.

\end{itemize}



Table \ref{tab:my-table1} contains the list of regression and classification tasks, detailing the chosen libraries and their respective versions used for implementation. We specifically aimed to select versions, outlined in Table \ref{tab:my-table1}, of Python and R that not only offer compatibility with the \textit{OS} (Ubuntu 20.04.6 LTS)  but also provide advanced memory management capabilities. For instance, Python version 3.8.10 was chosen for its implementation of shared memory segments for multiprocessing, which effectively reduces pickling costs between processes\footnote{\url{https://www.python.org/downloads/release/python-3810/}}. Similarly, we looked for compatible versions of R that offer enhanced memory management features, ensuring efficient handling of memory resources\footnote{\url{https://cran.r-project.org/doc/manuals/r-patched/NEWS.3.html}}. We employed well-established libraries  for Python and R, specifically from \textit{scikit-learn\footnote{\url{https://pypi.org/project/scikit-learn/}}} and the \textit{CRAN repository\footnote{\url{https://cran.r-project.org}}}, respectively. This is in line with several existing studies incorporating Python \cite{verdecchia2022data, hao2019machine} and R \cite{decan2015development, claes2015empirical} programming languages while dealing with ML tasks.



\vspace{-4mm}
\section{Datasets}
\label{datasets}
To assess and compare the energy consumption of machine learning algorithms, we conducted both classification and regression tasks using identical datasets for model training and inference. We strategically selected standard datasets from the UCI Machine Learning Repository and Kaggle datasets\footnote{\url{https://www.kaggle.com/datasets}}\cite{asuncion2007uci} based on predefined selection criteria, ensuring their relevance and applicability to our study. Additionally, we considered the potential impact of dataset size on energy consumption and incorporated two distinct datasets of varying sizes to address this variability. Below, we provide a brief description of the datasets used in this study, and Table \ref{tab:datasets-info} presents a summary of the key characteristics of these datasets.
\vspace{-3mm}

\subsection{Adult Dataset}
The dataset used in this study was derived from the 1994 census database. The primary task associated with this dataset involves classification, specifically determining whether an individual's annual income exceeds 50K. Dataset1, as referred to in the remainder of this paper, comprises 48,842 data points.

\subsection{Drug Review Dataset}
The dataset under consideration includes patient reviews pertaining to specific drugs, including information about the associated health conditions. It additionally provides a patient rating of the drug on a 10-star scale. It is referred to as Dataset2 for the rest of the paper and encompasses a total of 215,063 data points.

\subsection{New York City Taxi Trip Duration}
This dataset is taken from Kaggle\footnote{\url{https://www.kaggle.com/c/nyc-taxi-trip-duration/data}} and is based on the 2016 NYC Yellow Cab trip record data, made available in BigQuery on Google Cloud Platform. It includes detailed information about taxi trips, such as trip duration, pickup and drop-off locations, timestamps, and other relevant features. Throughout the rest of the paper, this dataset will be referred to as Dataset3. It comprises a total of 1,458,644 trip records, offering a comprehensive overview of taxi transportation in New York City during the specified period.

\section{Experimental Procedure}
\vspace{-2mm}
To systematically investigate the research questions, we conducted controlled experiments structured around training and inference phases of machine learning workflows. The following subsections detail the experimental design,setup used and methodology employed to assess energy consumption and runtime performance across Python and R implementations.

\label{experiments}

\vspace{-3mm}
\subsection{Experimental Design}
Our experiment comprises two phases of sub-experiments: one for model training and another for inferences, each involving distinct sets of independent and dependent variables. For the first independent variable (\textit{IV1}), we selected ten machine learning tasks. Additionally, we incorporated datasets with varying data points, detailed in Section \ref{datasets}. The number of data points for model training serves as the second independent variable (\textit{IV2}). This allows us to investigate the influence of dataset size on different machine learning tasks independently. In each sub-experiment, one independent variable is systematically altered while the other remains constant. The experimental results for the dependent variables, which is the energy consumption values(\textit{DV1}) and run-time(\textit{DV2}), are presented in Section \ref{results}. For training and making inferences with machine learning models in regression and classification tasks, we utilized implementations from the standard and reliable \textit{scikit-learn} and \textit{CRAN} repository in Python and R programming languages, respectively, as detailed in Table \ref{tab:my-table1}. A 60\%/20\%/20\% train/test/validation split was employed for model training and testing.

\begin{figure*}
    \centering
    \includegraphics[scale =0.38]{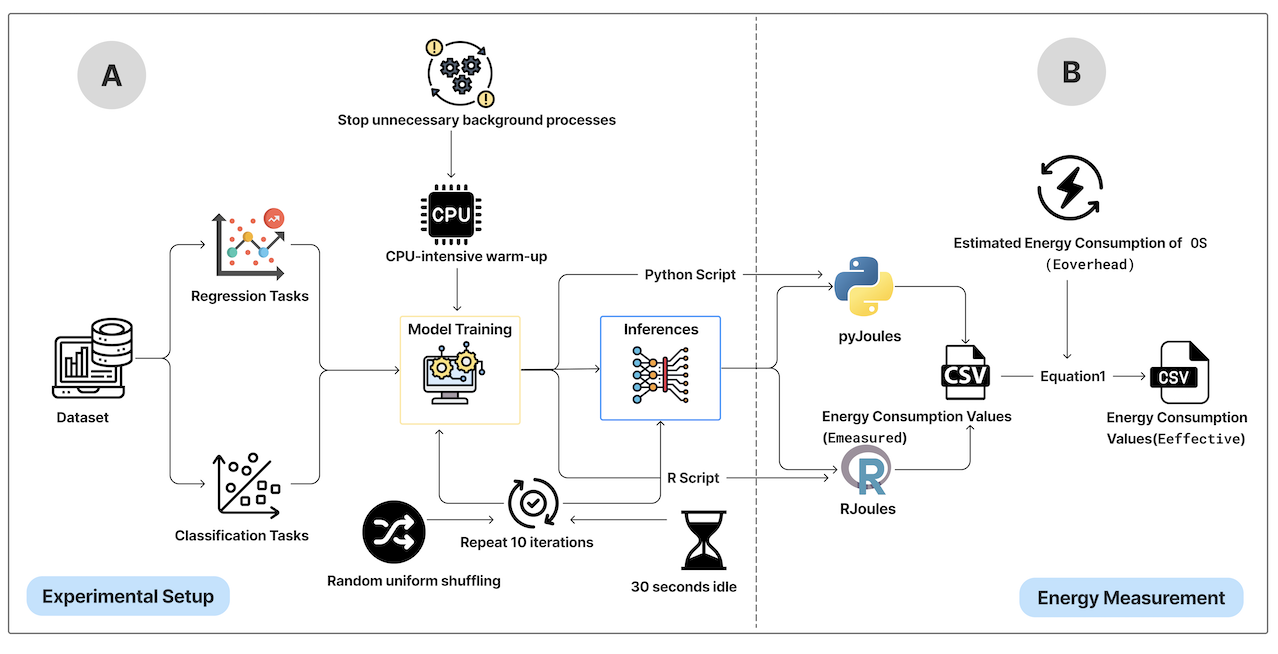}
    \caption{Workflow Diagram Describing the Experimental Procedure Followed in the Study}
    \label{fig:fig1}
\end{figure*}

\subsection{Experimental Setup}

The experiment was conducted on a Linux based System featuring Intel(R) Xeon(R) Gold 6226R CPU equipped with a frequency 2.90GHz, 16 cores, 128GB RAM. To measure the energy consumption of tasks executed in Python and R, we employed the tools  pyJoules and RJoules \cite{chattaraj2023rjoules}, respectively, built on Intel RAPL (Running Average Power Limit). RAPL functions as a power management interface on Intel x86 processors from the Sandy Bridge architecture and beyond, facilitating energy monitoring, power usage regulation, and the utilization of non-architectural Model Specific Registers (MSRs). Previous research within software contexts has demonstrated the dependability of the RAPL interface in gauging energy consumption \cite{david2010rapl,hahnel2012measuring, rotem2012power,shanbhag2023exploratory}. RAPL supports several physical domains of Intel processors, including Package(PKG), DRAM, Core and Uncore \cite{khan2018rapl}. 

\subsection{Procedure}
  The experimental procedure is outlined in Fig. \ref{fig:fig1}. At the beginning of the experiment, we systematically terminated all processes that were not essential for the operating system to function as followed by energy consumption related studies \cite{hindle2014greenminer,shanbhag2023exploratory,georgiou2022green} (Fig. \ref{fig:fig1}A). This precautionary measure was implemented to ensure minimal fluctuations in the readings, thereby mitigating the potential interference from background processes. To counter potential confounding factors due to increased temperatures in hardware components, a 5-second CPU-intensive warm-up preceded the experiment as followed by Verdecchia et al. \cite{verdecchia2022data} and illustrated in Fig. \ref{fig:fig1}A. To address potential noise and interference in energy consumption measurements, each task was repeated ten times, and the average energy measurement was recorded, following the methodology proposed by Georgiou et al. \cite{georgiou2022green} and depicted in Fig. \ref{fig:fig1}A. Moreover, a random uniform shuffling of sub-experiment executions was introduced to reduce noise impact further. For system stability, a 30-second idle period followed each iteration, aiming to mitigate the impact of power tail states \cite{bornholt2012model}. Based on this procedure, we assessed the energy consumption of various machine learning tasks (\textit{IV1}) implemented in both Python and R, as illustrated in Fig. \ref{fig:fig1}B.

  Additionally, to minimize the influence of background processes, both static and dynamic power unrelated to our experiment's focus, i.e., model training and inferences, were excluded. This process entailed quantifying the energy overhead by assessing the energy consumption of the operating system without any running scripts across multiple instances. Arafa et al. \cite{arafa2020verified} employed a similar approach to verify the energy consumption of PTX Instructions (Parallel Thread Execution). We rely on Arafa et al. \cite{arafa2020verified} and utilized Equation \ref{eqn:cpu} to compute the effective energy measurement, as outlined below.
  

\begin{equation}
\label{eqn:cpu}
    \boldsymbol{E_{effective} = E_{measured} - E_{overhead} }
\end{equation}


Here, $E_{\text{measured}}$ represents median energy consumption readings from pyJoules and RJoules, $E_{\text{overhead}}$ represents energy consumption without executing scripts, determined by measuring energy consumption through commenting out the scripts 10 times and obtaining the mean value before initiating the ML tasks, and $E_{\text{effective}}$ signifies energy consumption values for further analysis. Energy readings and run-time measurements from the experiments were stored in CSV files after each iteration, alongside algorithm execution times and corresponding energy data for each domain (Fig. \ref{fig:fig1}B). Furthermore, to evaluate any statistically significant variances and their scale, we present the outcomes of the Wilcoxon Signed Rank Test\cite{woolson2007wilcoxon} at a confidence level of 0.05. The scripts used for the study along with the detailed energy consumption and run-time values of the tasks for each trial, mean values, result analysis and standard deviation and coefficient of variation values are accessible at the following URL: {\url{https://anonymous.4open.science/r/RvsPython-CFE9/}.


\vspace{-3mm}
\section{Results}
\label{results}


In this section, we present the outcomes of empirical experiments, addressing the research questions. Tools such as pyJoules and RJoules [6] provide energy consumption data in micro-joules (1 micro-joule = 10\textsuperscript{-6} Joules). In our results, we present these measurements converted to Joules and rounded to two decimals for better readability. Our analysis of the results incorporates energy consumption values from the Package and DRAM domains. The underlying RAPL interface of both pyJoules and RJoules tools allows us to access energy consumption data for core and non-core components separately, but these are not explicitly reported as they are covered under the package(PKG) category. The results for Regression tasks are given in Table \ref{tab:my-table2}, and the results for Classification tasks are presented in Table \ref{tab:my-table3}. 
The energy consumption values in all the tables report mean values from 10 trials of each operation performed as described in Section \ref{experiments}. 


\subsection{Comparative Analysis of Python-scikit-learn and R Implementations}
\label{result1}
\textbf{A. Findings for RQ1.1: Comparing Python and R for Energy Efficiency and Run-Time Performance in Machine Learning Model Training for Regression Tasks.}

\begin{table*}[]
\centering
\caption{Mean Energy Consumption values(Joules) and Run Time(Seconds) for Regression Tasks-Using \textit{scikit-learn} Implementations of Python}

\label{tab:my-table2}
\resizebox{\textwidth}{!}{%
\begin{tabular}{|lllllllllllllll|}
\hline
\multicolumn{1}{|l|}{\multirow{3}{*}{\textbf{Machine Learning Algorithms(IV1)}}} & \multicolumn{7}{c|}{\textbf{Model Training Phase}} & \multicolumn{7}{c|}{\textbf{Inferences Phase}} \\ \cline{2-15} 
\multicolumn{1}{|l|}{} & \multicolumn{3}{l|}{\textbf{Python}} & \multicolumn{3}{l|}{\textbf{R}} & \multicolumn{1}{l|}{\textbf{p-value}} & \multicolumn{3}{l|}{\textbf{Python}} & \multicolumn{3}{l|}{\textbf{R}} & \textbf{p-value} \\ \cline{2-15} 
\multicolumn{1}{|l|}{} & \multicolumn{1}{l|}{\textbf{Run-Time}} & \multicolumn{1}{l|}{\textbf{Package}} & \multicolumn{1}{l|}{\textbf{DRAM}} & \multicolumn{1}{l|}{\textbf{Run-Time}} & \multicolumn{1}{l|}{\textbf{Package}} & \multicolumn{1}{l|}{\textbf{DRAM}} & \multicolumn{1}{l|}{\textbf{}} & \multicolumn{1}{l|}{\textbf{Run-Time}} & \multicolumn{1}{l|}{\textbf{Package}} & \multicolumn{1}{l|}{\textbf{DRAM}} & \multicolumn{1}{l|}{\textbf{Run-Time}} & \multicolumn{1}{l|}{\textbf{Package}} & \multicolumn{1}{l|}{\textbf{DRAM}} & \textbf{} \\ \hline
\multicolumn{15}{|c|}{\textbf{Dataset1(IV2)}} \\ \hline
\multicolumn{1}{|l|}{\textbf{Decision Tree}} & \multicolumn{1}{l|}{0.08} & \multicolumn{1}{l|}{5} & \multicolumn{1}{l|}{5} & \multicolumn{1}{l|}{0.02} & \multicolumn{1}{l|}{2.15} & \multicolumn{1}{l|}{0.22} & \multicolumn{1}{l|}{0.002} & \multicolumn{1}{l|}{0.008} & \multicolumn{1}{l|}{0.41} & \multicolumn{1}{l|}{0.06} & \multicolumn{1}{l|}{0.003} & \multicolumn{1}{l|}{0.37} & \multicolumn{1}{l|}{0.03} & 0.008 \\ \hline
\multicolumn{1}{|l|}{\textbf{Gaussian Regression}} & \multicolumn{1}{l|}{31.46} & \multicolumn{1}{l|}{4483.33} & \multicolumn{1}{l|}{365.55} & \multicolumn{1}{l|}{0.38} & \multicolumn{1}{l|}{36.92} & \multicolumn{1}{l|}{3.64} & \multicolumn{1}{l|}{0.002} & \multicolumn{1}{l|}{8.234} & \multicolumn{1}{l|}{1504.26} & \multicolumn{1}{l|}{123.24} & \multicolumn{1}{l|}{0.023} & \multicolumn{1}{l|}{2.39} & \multicolumn{1}{l|}{0.24} & 0.002 \\ \hline
\multicolumn{1}{|l|}{\textbf{Linear Regression}} & \multicolumn{1}{l|}{0.019} & \multicolumn{1}{l|}{1.07} & \multicolumn{1}{l|}{0.16} & \multicolumn{1}{l|}{0.18} & \multicolumn{1}{l|}{17.59} & \multicolumn{1}{l|}{1.71} & \multicolumn{1}{l|}{0.002} & \multicolumn{1}{l|}{0.009} & \multicolumn{1}{l|}{0.59} & \multicolumn{1}{l|}{0.07} & \multicolumn{1}{l|}{0.18} & \multicolumn{1}{l|}{17.5} & \multicolumn{1}{l|}{1.75} & 0.002 \\ \hline
\multicolumn{1}{|l|}{\textbf{Neural Network Regression}} & \multicolumn{1}{l|}{11.624} & \multicolumn{1}{l|}{3501.85} & \multicolumn{1}{l|}{113.12} & \multicolumn{1}{l|}{0.58} & \multicolumn{1}{l|}{56.79} & \multicolumn{1}{l|}{5.88} & \multicolumn{1}{l|}{0.002} & \multicolumn{1}{l|}{0.03} & \multicolumn{1}{l|}{3.53} & \multicolumn{1}{l|}{0.3} & \multicolumn{1}{l|}{0.002} & \multicolumn{1}{l|}{0.32} & \multicolumn{1}{l|}{0.03} & 0.002 \\ \hline
\multicolumn{1}{|l|}{\textbf{Support Vector Machine}} & \multicolumn{1}{l|}{9.01} & \multicolumn{1}{l|}{885.85} & \multicolumn{1}{l|}{76.54} & \multicolumn{1}{l|}{19.69} & \multicolumn{1}{l|}{1923} & \multicolumn{1}{l|}{175} & \multicolumn{1}{l|}{0.002} & \multicolumn{1}{l|}{2.399} & \multicolumn{1}{l|}{234.16} & \multicolumn{1}{l|}{19.83} & \multicolumn{1}{l|}{0.003} & \multicolumn{1}{l|}{0.37} & \multicolumn{1}{l|}{0.03} & 0.002 \\ \hline
\multicolumn{15}{|c|}{\textbf{Dataset2(IV2)}} \\ \hline
\multicolumn{1}{|l|}{\textbf{Decision Tree}} & \multicolumn{1}{l|}{0.99} & \multicolumn{1}{l|}{17.07} & \multicolumn{1}{l|}{9.02} & \multicolumn{1}{l|}{0.143} & \multicolumn{1}{l|}{10.65} & \multicolumn{1}{l|}{1.36} & \multicolumn{1}{l|}{0.002} & \multicolumn{1}{l|}{0.013} & \multicolumn{1}{l|}{1.23} & \multicolumn{1}{l|}{0.66} & \multicolumn{1}{l|}{0.009} & \multicolumn{1}{l|}{0.73} & \multicolumn{1}{l|}{0.11} & 0.002 \\ \hline
\multicolumn{1}{|l|}{\textbf{Gaussian Regression}} & \multicolumn{1}{l|}{632.16} & \multicolumn{1}{l|}{69838.74} & \multicolumn{1}{l|}{982.62} & \multicolumn{1}{l|}{26.398} & \multicolumn{1}{l|}{7124.58} & \multicolumn{1}{l|}{573.21} & \multicolumn{1}{l|}{0.002} & \multicolumn{1}{l|}{147.47} & \multicolumn{1}{l|}{9543.32} & \multicolumn{1}{l|}{1123.58} & \multicolumn{1}{l|}{3.291} & \multicolumn{1}{l|}{322.63} & \multicolumn{1}{l|}{31.2} & 0.002 \\ \hline
\multicolumn{1}{|l|}{\textbf{Linear Regression}} & \multicolumn{1}{l|}{0.019} & \multicolumn{1}{l|}{1021.08} & \multicolumn{1}{l|}{57.16} & \multicolumn{1}{l|}{20.997} & \multicolumn{1}{l|}{38485.03} & \multicolumn{1}{l|}{2533.54} & \multicolumn{1}{l|}{0.002} & \multicolumn{1}{l|}{0.005} & \multicolumn{1}{l|}{34.67} & \multicolumn{1}{l|}{5.73} & \multicolumn{1}{l|}{19.727} & \multicolumn{1}{l|}{501.42} & \multicolumn{1}{l|}{39.7} & \textbf{0.48} \\ \hline
\multicolumn{1}{|l|}{\textbf{Neural Network Regression}} & \multicolumn{1}{l|}{5.09} & \multicolumn{1}{l|}{7087.74} & \multicolumn{1}{l|}{854.98} & \multicolumn{1}{l|}{16.32} & \multicolumn{1}{l|}{356.38} & \multicolumn{1}{l|}{46.34} & \multicolumn{1}{l|}{0.002} & \multicolumn{1}{l|}{0.04} & \multicolumn{1}{l|}{8.49} & \multicolumn{1}{l|}{3.39} & \multicolumn{1}{l|}{0.008} & \multicolumn{1}{l|}{1.45} & \multicolumn{1}{l|}{0.83} & 0.002 \\ \hline
\multicolumn{1}{|l|}{\textbf{Support Vector Machine}} & \multicolumn{1}{l|}{107.98} & \multicolumn{1}{l|}{4702.77} & \multicolumn{1}{l|}{5913.24} & \multicolumn{1}{l|}{9.366} & \multicolumn{1}{l|}{55962.95} & \multicolumn{1}{l|}{4545.79} & \multicolumn{1}{l|}{\textbf{0.14}} & \multicolumn{1}{l|}{41.52} & \multicolumn{1}{l|}{3426.23} & \multicolumn{1}{l|}{582.33} & \multicolumn{1}{l|}{2.115} & \multicolumn{1}{l|}{28.76} & \multicolumn{1}{l|}{2.73} & 0.002 \\ \hline
\multicolumn{15}{|c|}{\textbf{Dataset3(IV2)}} \\ \hline
\multicolumn{1}{|l|}{\textbf{Decision Tree}} & \multicolumn{1}{l|}{3.5} & \multicolumn{1}{l|}{28.68} & \multicolumn{1}{l|}{3.84} & \multicolumn{1}{l|}{2.7} & \multicolumn{1}{l|}{214.3} & \multicolumn{1}{l|}{27.2} & \multicolumn{1}{l|}{0.002} & \multicolumn{1}{l|}{0.06} & \multicolumn{1}{l|}{3.26} & \multicolumn{1}{l|}{0.68} & \multicolumn{1}{l|}{0.32} & \multicolumn{1}{l|}{27.4} & \multicolumn{1}{l|}{3.1} & 0.002 \\ \hline
\multicolumn{1}{|l|}{\textbf{Gaussian Regression}} & \multicolumn{1}{l|}{729.2} & \multicolumn{1}{l|}{89666.6} & \multicolumn{1}{l|}{7311} & \multicolumn{1}{l|}{26.421} & \multicolumn{1}{l|}{21444.22} & \multicolumn{1}{l|}{1756.62} & \multicolumn{1}{l|}{0.002} & \multicolumn{1}{l|}{164.68} & \multicolumn{1}{l|}{190866.4} & \multicolumn{1}{l|}{22471.6} & \multicolumn{1}{l|}{122.43} & \multicolumn{1}{l|}{12001.38} & \multicolumn{1}{l|}{1160.65} & 0.002 \\ \hline
\multicolumn{1}{|l|}{\textbf{Linear Regression}} & \multicolumn{1}{l|}{0.09} & \multicolumn{1}{l|}{4.85} & \multicolumn{1}{l|}{0.7} & \multicolumn{1}{l|}{20.999} & \multicolumn{1}{l|}{43485.03} & \multicolumn{1}{l|}{3534.59} & \multicolumn{1}{l|}{0.002} & \multicolumn{1}{l|}{0.02} & \multicolumn{1}{l|}{1.04} & \multicolumn{1}{l|}{0.16} & \multicolumn{1}{l|}{0.63} & \multicolumn{1}{l|}{541.3} & \multicolumn{1}{l|}{43.5} & 0.002 \\ \hline
\multicolumn{1}{|l|}{\textbf{Neural Network Regression}} & \multicolumn{1}{l|}{22.91} & \multicolumn{1}{l|}{6934.02} & \multicolumn{1}{l|}{237.09} & \multicolumn{1}{l|}{92.5} & \multicolumn{1}{l|}{2685.9} & \multicolumn{1}{l|}{421.5} & \multicolumn{1}{l|}{0.002} & \multicolumn{1}{l|}{0.18} & \multicolumn{1}{l|}{11.21} & \multicolumn{1}{l|}{1.59} & \multicolumn{1}{l|}{732.651} & \multicolumn{1}{l|}{7276.54} & \multicolumn{1}{l|}{423119.21} & 0.002 \\ \hline
\multicolumn{1}{|l|}{\textbf{Support Vector Machine}} & \multicolumn{1}{l|}{485.93} & \multicolumn{1}{l|}{47772.06} & \multicolumn{1}{l|}{4559.84} & \multicolumn{1}{l|}{46.8} & \multicolumn{1}{l|}{62420.8} & \multicolumn{1}{l|}{4885.5} & \multicolumn{1}{l|}{0.002} & \multicolumn{1}{l|}{186.86} & \multicolumn{1}{l|}{18363.89} & \multicolumn{1}{l|}{1660.99} & \multicolumn{1}{l|}{78.5976} & \multicolumn{1}{l|}{470229.571} & \multicolumn{1}{l|}{36848.38} & 0.002 \\ \hline
\end{tabular}%
}
\end{table*}

Through the first research question (RQ1), we aim to determine which programming language (Python or R) is more energy and runtime-efficient for training machine learning models across multiple regression tasks.
Analyzing the results from Dataset1, R generally outperforms Python in both energy and run-time efficiency for most models, except for Support Vector Machine (SVM) Regression and Linear Regression, as provided in Table \ref{tab:my-table2}. Specifically, in Gaussian Regression, R is 99.16\% more energy-efficient, and in Neural Network Regression, it shows 98.2\% greater energy efficiency. However, Python consumes 45.9\% less energy in SVM Regression. In terms of run-time performance, R demonstrates a significant speed advantage in Decision Tree, Gaussian Regression, and Neural Network Regression (4x, 82x, and 20x faster, respectively). Meanwhile, Python is 9x faster in Linear Regression and 2x faster in SVM Regression. In Dataset2, R continues to show better energy efficiency in Decision Tree, Gaussian Regression, and Neural Network Regression, with improvements of 54\%, 89.1\%, and 94.9\%, respectively. However, Python is more energy-efficient in Linear Regression and SVM Regression, showing a 90.2\% and 65.2\% energy advantage, respectively. For run-time performance, R retains its advantage in Decision Tree, Gaussian Regression, and Neural Network Regression, while Python is faster in Linear Regression, with a notable 1105x speed increase. In Dataset3, Python shows 86.6\% greater energy efficiency in Decision Tree Regression, though it is 1.3x slower. For Gaussian Regression, R consumes 31.5\% less energy and is 27.6x faster than Python. In Neural Network Regression, R consumes 75\% more energy, but Python is 4x faster. For SVM Regression, Python consumes 48.3\% less energy, but R is faster. Overall, while R is more energy-efficient for Gaussian Regression, Python demonstrates superior energy efficiency for Neural Network and SVM Regression, especially for larger datasets like Dataset3, as given in Table\ref{tab:my-table2}.

 
 
 These findings suggest that the choice of programming language can have a substantial impact on energy efficiency, emphasizing the need to consider specific algorithmic requirements and constraints when selecting between Python and R for regression tasks.




\begin{figure}
    \centering
    \includegraphics[scale=0.43]{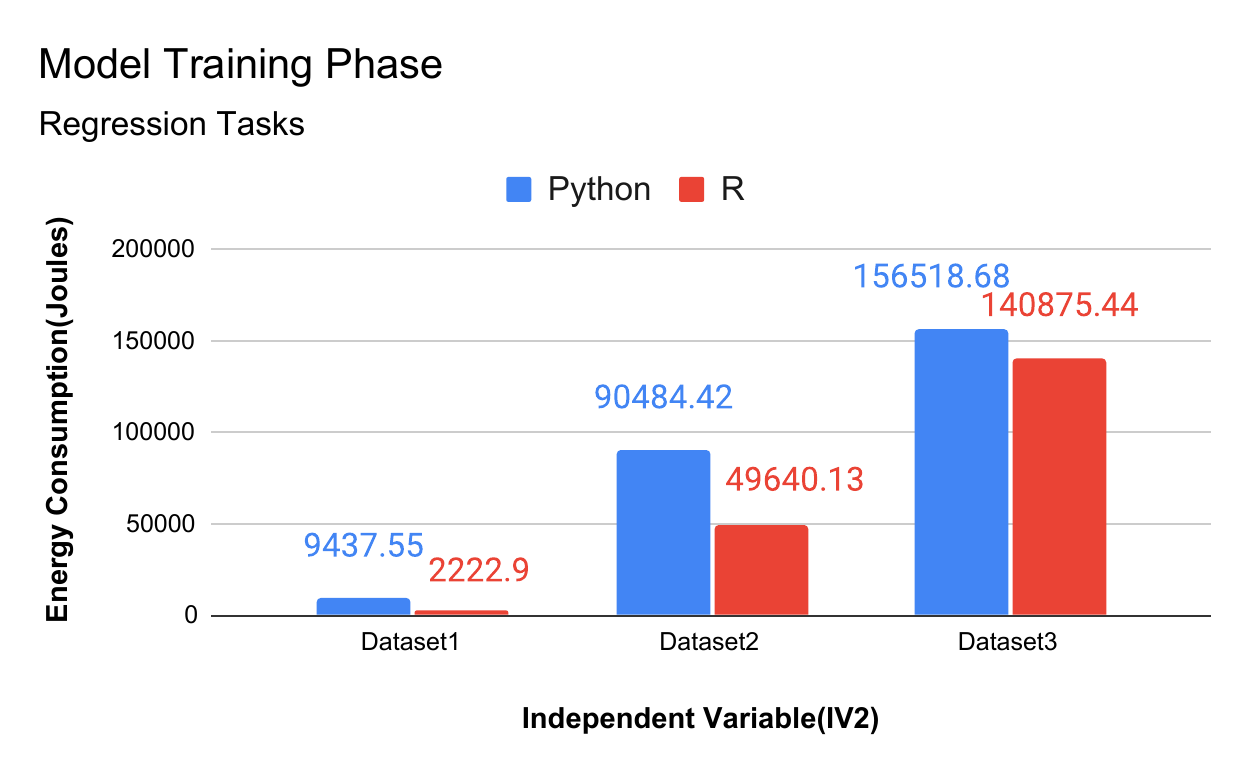}
    \caption{Cumulative Median Energy Consumption of Regression Tasks for Model Training Phase}
    \label{fig:regression1}
\end{figure}

\textbf{B. Findings for RQ1.2: Comparing Python and R for Energy Efficiency and Run-Time Performance in Machine Learning Model Inferences for Regression Tasks.}
\begin{figure}
    \centering
    
    \includegraphics[scale=0.45]{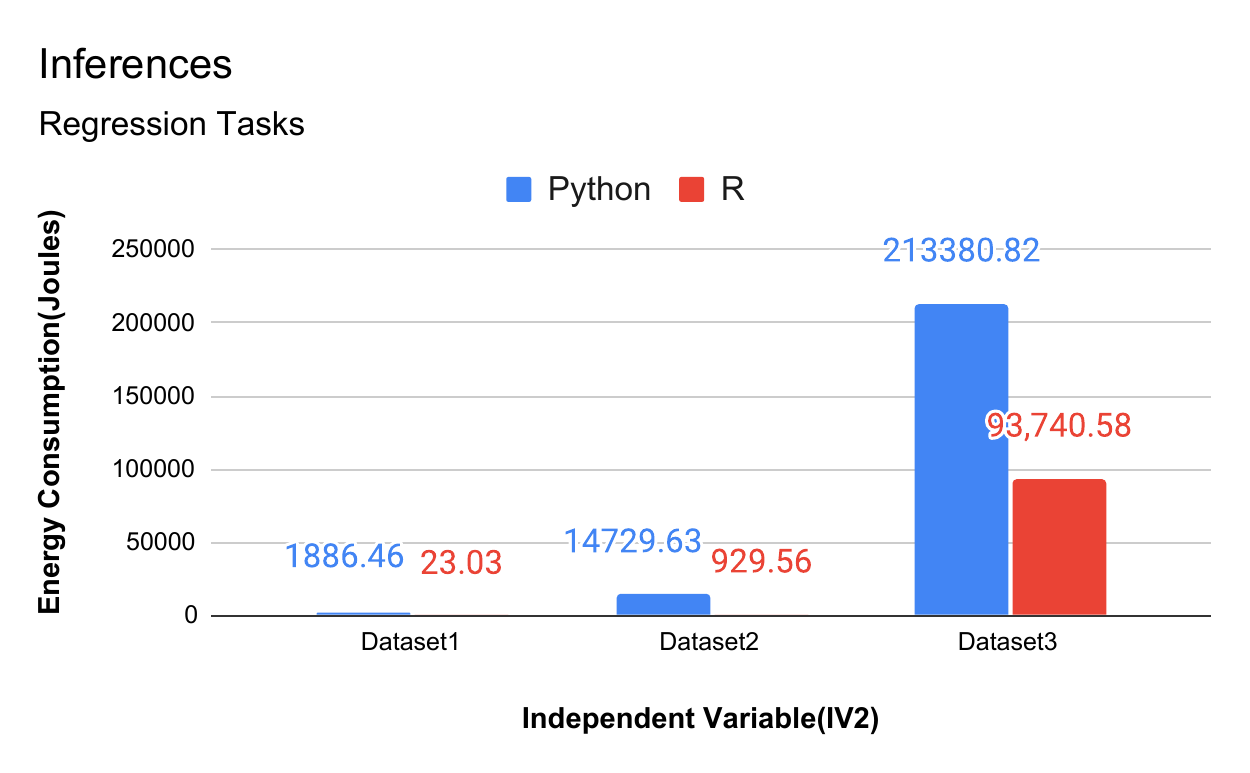}
    \caption{Cumulative Median Energy Consumption of Regression Tasks for Inferences Phase}
    \label{fig:regression2}
\end{figure}

This research question aims to determine whether Python or R exhibits better energy efficiency and run-time performance in performing inferences across various regression tasks. To address this, initially we assigned Dataset1 as the independent variable (IV2). Analysis from Table \ref{tab:my-table2} highlights R's superior energy efficiency across all tasks except for Linear regression inferences. Notably, in Gaussian Regression and SVM inferences, R demonstrates 99.8\% more energy efficiency than Python, while Python provides 96.6\% more energy-efficiency than R in the case of Linear Regression. Shifting to Dataset2, R continues to showcase higher energy efficiency across all algorithms except for Linear Regression. Python exhibits 92.5\% more energy efficiency than R for Linear Regression. When considering the minimum and maximum variations, R displays a minimum decrease of 55.5\% in energy consumption (for Decision Tree) and a maximum decrease of 99.2\% (for SVM). In Dataset3 inferences, R shows greater energy efficiency for Decision Tree and Neural Network Regression, with reductions of 97.6\% and 98.5\%, respectively, compared to Python. However, in Gaussian Regression, Python is 96.3\% more energy-efficient, while for Linear Regression and SVM, Python is 98.4\% and 92.0\% more energy-efficient than R, respectively.


Similar to the model training phase of regression tasks, the runtime performance shows a similar trend to energy efficiency across all 3 datasets. R demonstrates runtime efficiency across all regression tasks during inference, except for Linear Regression. Notably, Python outperforms R by 3945 times during Linear Regression inference in Dataset2.

The cumulative energy consumption values while training machine learning models for different regression tasks over the three selected datasets are depicted in Fig. \ref{fig:regression1}. The results reveal notable discrepancies in cumulative energy consumption between Python and R across the datasets. Cumulatively, R is observed to be \textbf{76.4\%} and \textbf{45.1\%} more energy-efficient than Python in \textbf{Dataset1} and \textbf{Dataset2}, respectively. However, in \textbf{Dataset3}, Python is \textbf{10\% more energy-efficient} than R.

\begin{tcolorbox}[colback=blue!10!white, colframe=blue!50!black, title=Key Insights from RQ1.1 and RQ1.2]
\small
\begin{itemize}
    \item In the training phase of regression tasks, R is more energy and run-time efficient than Python for Decision Tree, Gaussian Regression, and Neural Network Regression. 


\item During the inference phase of regression tasks, R exhibits superior energy efficiency across all algorithms except for Linear Regression.

\item Regarding SVM, R displays higher energy and run-time efficiency during model training, whereas Python is more energy efficient during the inference phase.

\item R outperforms Python by up to 99.16\% in energy efficiency during Neural Network Regression model training and achieves up to 99.8\% higher efficiency in the inference phase for Gaussian Regression and SVM.

\item  In Dataset3, R shows energy savings, being 11.11\% more energy-efficient than Python in Linear Regression during the model training phase and 127.67\% more energy-efficient during the inference phase.

\end{itemize}

\end{tcolorbox}

Similarly, during inferences, as shown in Fig. \ref{fig:regression2}, R demonstrates \textbf{98.8\%} and \textbf{93.7\%} more energy efficiency for \textbf{Dataset1} and \textbf{Dataset2}, respectively. In contrast, in \textbf{Dataset3}, Python is \textbf{56.5\% more energy-efficient} than R.

An essential finding is that as the number of data points increases from approximately 49k to 0.2M to 1.4M, Python's energy consumption increases by \textbf{9.5 times} between \textbf{Dataset1} and \textbf{Dataset2}, while for R, it increases by \textbf{22.3 times} during model training. In \textbf{Dataset3}, Python’s energy consumption increases by an additional \textbf{1.7 times}, while R’s energy consumption increases by \textbf{2.8 times} compared to \textbf{Dataset2}.

Specifically, for \textbf{Linear Regression}, Python exhibits an \textbf{829 times} increase and R shows a \textbf{571 times} increase in energy consumption from \textbf{Dataset1} to \textbf{Dataset2}. In the inference phase, Python's energy consumption increases \textbf{7.8 times}, while for R, it increases \textbf{40 times} between \textbf{Dataset1} and \textbf{Dataset2}. As we move to  \textbf{Dataset3}, Python's inference energy consumption increases by an additional \textbf{14.5 times}, while R’s energy consumption increases by \textbf{100 times} compared to \textbf{Dataset2}. These observations suggest that as the number of data points increases, the energy consumption of R language implementation for Regression Tasks increases more rapidly than that of Python. However, further studies are required to validate this observation, considering the potential impact of the dataset type.

\begin{figure}
    \centering
    
    \includegraphics[scale=0.46]{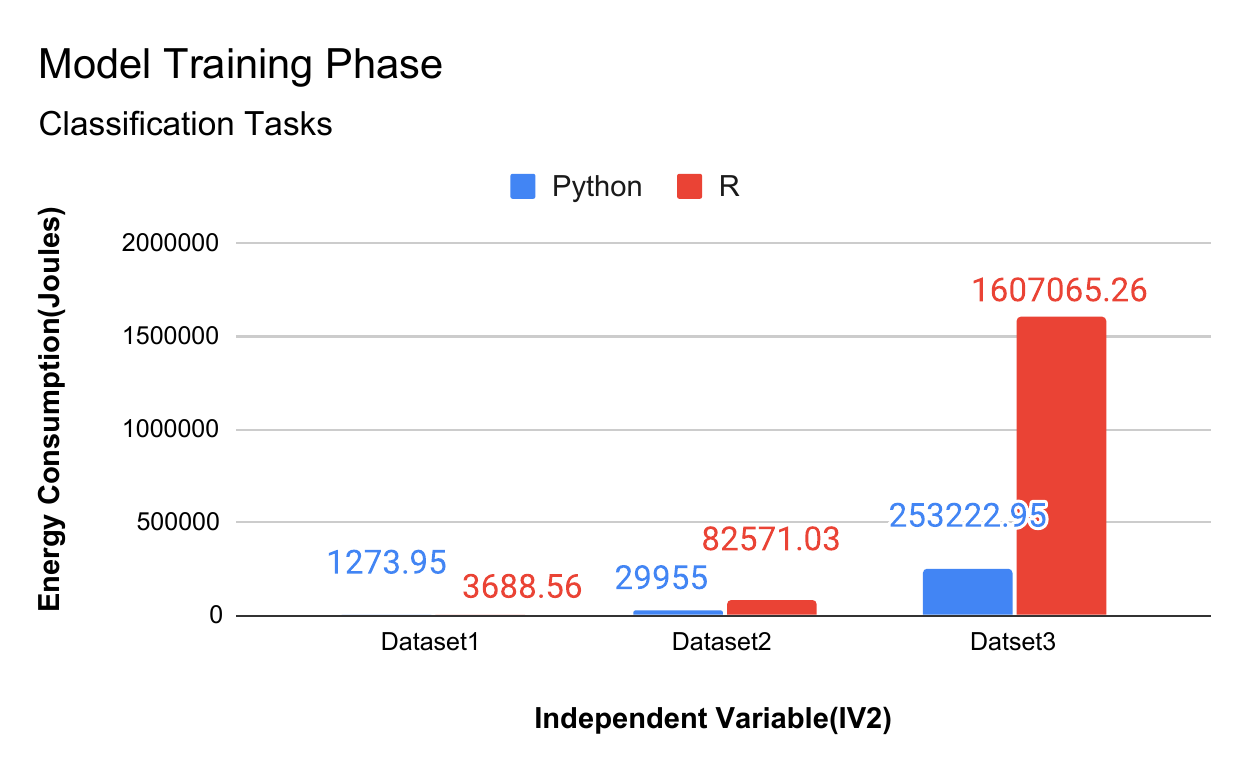}
    \caption{Cumulative Median Energy Consumption of Classification Tasks for Model Training Phase}
    \label{fig:classification1}
\end{figure}

\begin{table*}[]
\centering
\caption{Mean Energy Consumption values (Joules) and Run Time (Seconds) for Classification Tasks-Using \textit{scikit-learn} Implementations of Python}
\vspace{2mm}
\label{tab:my-table3}
\resizebox{\textwidth}{!}{%
\begin{tabular}{|lllllllllllllll|}
\hline
\multicolumn{1}{|l|}{\multirow{3}{*}{\textbf{Machine Learning Algorithms(IV1)}}} & \multicolumn{7}{c|}{\textbf{Model Training Phase}} & \multicolumn{7}{c|}{\textbf{Inferences Phase}} \\ \cline{2-15} 
\multicolumn{1}{|l|}{} & \multicolumn{3}{l|}{\textbf{Python}} & \multicolumn{3}{l|}{\textbf{R}} & \multicolumn{1}{l|}{\multirow{2}{*}{\textbf{p-value}}} & \multicolumn{3}{l|}{Python} & \multicolumn{3}{l|}{R} & \multirow{2}{*}{\textbf{p-value}} \\ \cline{2-7} \cline{9-14}
\multicolumn{1}{|l|}{} & \multicolumn{1}{l|}{\textbf{Run-Time}} & \multicolumn{1}{l|}{\textbf{Package}} & \multicolumn{1}{l|}{\textbf{DRAM}} & \multicolumn{1}{l|}{\textbf{Run-Time}} & \multicolumn{1}{l|}{\textbf{Package}} & \multicolumn{1}{l|}{\textbf{DRAM}} & \multicolumn{1}{l|}{} & \multicolumn{1}{l|}{Run-Time} & \multicolumn{1}{l|}{Package} & \multicolumn{1}{l|}{DRAM} & \multicolumn{1}{l|}{Run-Time} & \multicolumn{1}{l|}{Package} & \multicolumn{1}{l|}{DRAM} &  \\ \hline
\multicolumn{15}{|c|}{\textbf{Dataset1(IV2)}} \\ \hline
\multicolumn{1}{|l|}{\textbf{Decision Tree}} & \multicolumn{1}{l|}{5.285} & \multicolumn{1}{l|}{513.95} & \multicolumn{1}{l|}{48.08} & \multicolumn{1}{l|}{0.476} & \multicolumn{1}{l|}{46.34} & \multicolumn{1}{l|}{4.3} & \multicolumn{1}{l|}{0.002} & \multicolumn{1}{l|}{1.904} & \multicolumn{1}{l|}{182.38} & \multicolumn{1}{l|}{16.87} & \multicolumn{1}{l|}{0.022} & \multicolumn{1}{l|}{2.25} & \multicolumn{1}{l|}{0.23} & 0.002 \\ \hline
\multicolumn{1}{|l|}{\textbf{Gaussian Naive Bayes}} & \multicolumn{1}{l|}{0.011} & \multicolumn{1}{l|}{0.63} & \multicolumn{1}{l|}{0.09} & \multicolumn{1}{l|}{0.019} & \multicolumn{1}{l|}{1.99} & \multicolumn{1}{l|}{0.19} & \multicolumn{1}{l|}{0.002} & \multicolumn{1}{l|}{0.006} & \multicolumn{1}{l|}{0.3} & \multicolumn{1}{l|}{0.05} & \multicolumn{1}{l|}{0.001} & \multicolumn{1}{l|}{0.21} & \multicolumn{1}{l|}{0.02} & 0.002 \\ \hline
\multicolumn{1}{|l|}{\textbf{Logistic Regression}} & \multicolumn{1}{l|}{0.039} & \multicolumn{1}{l|}{4.74} & \multicolumn{1}{l|}{0.33} & \multicolumn{1}{l|}{2.345} & \multicolumn{1}{l|}{226.63} & \multicolumn{1}{l|}{21.84} & \multicolumn{1}{l|}{0.002} & \multicolumn{1}{l|}{0.009} & \multicolumn{1}{l|}{0.71} & \multicolumn{1}{l|}{0.08} & \multicolumn{1}{l|}{0.03} & \multicolumn{1}{l|}{3.09} & \multicolumn{1}{l|}{0.32} & 0.002 \\ \hline
\multicolumn{1}{|l|}{\textbf{Random Forest}} & \multicolumn{1}{l|}{1.395} & \multicolumn{1}{l|}{129.82} & \multicolumn{1}{l|}{12.43} & \multicolumn{1}{l|}{8.173} & \multicolumn{1}{l|}{794.97} & \multicolumn{1}{l|}{66.18} & \multicolumn{1}{l|}{0.002} & \multicolumn{1}{l|}{0.055} & \multicolumn{1}{l|}{3.41} & \multicolumn{1}{l|}{0.46} & \multicolumn{1}{l|}{0.406} & \multicolumn{1}{l|}{39.15} & \multicolumn{1}{l|}{3.33} & 0.002 \\ \hline
\multicolumn{1}{|l|}{\textbf{Support Vector Machine}} & \multicolumn{1}{l|}{5.295} & \multicolumn{1}{l|}{514.83} & \multicolumn{1}{l|}{48.05} & \multicolumn{1}{l|}{23.48} & \multicolumn{1}{l|}{2343.45} & \multicolumn{1}{l|}{182.67} & \multicolumn{1}{l|}{0.002} & \multicolumn{1}{l|}{1.916} & \multicolumn{1}{l|}{182.44} & \multicolumn{1}{l|}{16.88} & \multicolumn{1}{l|}{7.242} & \multicolumn{1}{l|}{725.12} & \multicolumn{1}{l|}{56.01} & 0.002 \\ \hline
\multicolumn{15}{|c|}{\textbf{Dataset2(IV2)}} \\ \hline
\multicolumn{1}{|l|}{\textbf{Decision Tree}} & \multicolumn{1}{l|}{115.451} & \multicolumn{1}{l|}{11303.81} & \multicolumn{1}{l|}{6780.85} & \multicolumn{1}{l|}{0.596} & \multicolumn{1}{l|}{57.59} & \multicolumn{1}{l|}{34.1} & \multicolumn{1}{l|}{0.002} & \multicolumn{1}{l|}{43.234} & \multicolumn{1}{l|}{4218.79} & \multicolumn{1}{l|}{2527.05} & \multicolumn{1}{l|}{0.064} & \multicolumn{1}{l|}{6.5} & \multicolumn{1}{l|}{3.72} & 0.002 \\ \hline
\multicolumn{1}{|l|}{\textbf{Gaussian Naive Bayes}} & \multicolumn{1}{l|}{0.016} & \multicolumn{1}{l|}{0.84} & \multicolumn{1}{l|}{0.49} & \multicolumn{1}{l|}{0.097} & \multicolumn{1}{l|}{9.6} & \multicolumn{1}{l|}{5.66} & \multicolumn{1}{l|}{0.002} & \multicolumn{1}{l|}{0.007} & \multicolumn{1}{l|}{0.37} & \multicolumn{1}{l|}{0.22} & \multicolumn{1}{l|}{0.004} & \multicolumn{1}{l|}{0.49} & \multicolumn{1}{l|}{0.29} & 0.046 \\ \hline
\multicolumn{1}{|l|}{\textbf{Logistic Regression}} & \multicolumn{1}{l|}{0.062} & \multicolumn{1}{l|}{6.93} & \multicolumn{1}{l|}{4.44} & \multicolumn{1}{l|}{24.34} & \multicolumn{1}{l|}{528.34} & \multicolumn{1}{l|}{237} & \multicolumn{1}{l|}{0.002} & \multicolumn{1}{l|}{0.006} & \multicolumn{1}{l|}{0.16} & \multicolumn{1}{l|}{0.17} & \multicolumn{1}{l|}{0.086} & \multicolumn{1}{l|}{5.02} & \multicolumn{1}{l|}{1.3} & 0.002 \\ \hline
\multicolumn{1}{|l|}{\textbf{Random Forest}} & \multicolumn{1}{l|}{5.716} & \multicolumn{1}{l|}{247.34} & \multicolumn{1}{l|}{301.39} & \multicolumn{1}{l|}{67.23} & \multicolumn{1}{l|}{1037.92} & \multicolumn{1}{l|}{169.45} & \multicolumn{1}{l|}{0.002} & \multicolumn{1}{l|}{0.145} & \multicolumn{1}{l|}{4.71} & \multicolumn{1}{l|}{5.44} & \multicolumn{1}{l|}{3.78} & \multicolumn{1}{l|}{191.65} & \multicolumn{1}{l|}{22.67} & 0.002 \\ \hline
\multicolumn{1}{|l|}{\textbf{Support Vector Machine}} & \multicolumn{1}{l|}{108.267} & \multicolumn{1}{l|}{5032.91} & \multicolumn{1}{l|}{6276} & \multicolumn{1}{l|}{4.041} & \multicolumn{1}{l|}{76371.35} & \multicolumn{1}{l|}{4120.02} & \multicolumn{1}{l|}{\textbf{0.19}} & \multicolumn{1}{l|}{43.191} & \multicolumn{1}{l|}{1880.27} & \multicolumn{1}{l|}{2341.05} & \multicolumn{1}{l|}{4.04} & \multicolumn{1}{l|}{14151.96} & \multicolumn{1}{l|}{1132.35} & 0.002 \\ \hline
\multicolumn{15}{|c|}{\textbf{Dataset3(IV2)}} \\ \hline
\multicolumn{1}{|l|}{\textbf{Decision Tree}} & \multicolumn{1}{l|}{2309.013} & \multicolumn{1}{l|}{113038.09} & \multicolumn{1}{l|}{10504.14} & \multicolumn{1}{l|}{9.53} & \multicolumn{1}{l|}{921.47} & \multicolumn{1}{l|}{86.31} & \multicolumn{1}{l|}{0.002} & \multicolumn{1}{l|}{864.672} & \multicolumn{1}{l|}{42187.88} & \multicolumn{1}{l|}{3824.54} & \multicolumn{1}{l|}{1.028} & \multicolumn{1}{l|}{101.53} & \multicolumn{1}{l|}{10.18} & 0.002 \\ \hline
\multicolumn{1}{|l|}{\textbf{Gaussian Naive Bayes}} & \multicolumn{1}{l|}{0.311} & \multicolumn{1}{l|}{8.39} & \multicolumn{1}{l|}{1.22} & \multicolumn{1}{l|}{1.548} & \multicolumn{1}{l|}{153.56} & \multicolumn{1}{l|}{14.69} & \multicolumn{1}{l|}{0.002} & \multicolumn{1}{l|}{0.144} & \multicolumn{1}{l|}{3.73} & \multicolumn{1}{l|}{0.58} & \multicolumn{1}{l|}{0.064} & \multicolumn{1}{l|}{7.8} & \multicolumn{1}{l|}{0.71} & 0.002 \\ \hline
\multicolumn{1}{|l|}{\textbf{Logistic Regression}} & \multicolumn{1}{l|}{1.233} & \multicolumn{1}{l|}{69.28} & \multicolumn{1}{l|}{5.42} & \multicolumn{1}{l|}{105.78} & \multicolumn{1}{l|}{45267.23} & \multicolumn{1}{l|}{1345.78} & \multicolumn{1}{l|}{0.002} & \multicolumn{1}{l|}{0.126} & \multicolumn{1}{l|}{3.31} & \multicolumn{1}{l|}{0.51} & \multicolumn{1}{l|}{15.89} & \multicolumn{1}{l|}{56.55} & \multicolumn{1}{l|}{267.21} & 0.002 \\ \hline
\multicolumn{1}{|l|}{\textbf{Random Forest}} & \multicolumn{1}{l|}{114.319} & \multicolumn{1}{l|}{5487.31} & \multicolumn{1}{l|}{512.46} & \multicolumn{1}{l|}{195.43} & \multicolumn{1}{l|}{42356} & \multicolumn{1}{l|}{1876.11} & \multicolumn{1}{l|}{0.002} & \multicolumn{1}{l|}{2.902} & \multicolumn{1}{l|}{101.55} & \multicolumn{1}{l|}{12.63} & \multicolumn{1}{l|}{121.72} & \multicolumn{1}{l|}{1099.34} & \multicolumn{1}{l|}{654.12} & 0.002 \\ \hline
\multicolumn{1}{|l|}{\textbf{Support Vector Machine}} & \multicolumn{1}{l|}{2300.759} & \multicolumn{1}{l|}{113089.16} & \multicolumn{1}{l|}{10507.48} & \multicolumn{1}{l|}{233.582} & \multicolumn{1}{l|}{1396581.56} & \multicolumn{1}{l|}{114462.55} & \multicolumn{1}{l|}{0.002} & \multicolumn{1}{l|}{863.819} & \multicolumn{1}{l|}{42213.26} & \multicolumn{1}{l|}{3820.44} & \multicolumn{1}{l|}{64.655} & \multicolumn{1}{l|}{386431.3} & \multicolumn{1}{l|}{30540.43} & 0.002 \\ \hline
\end{tabular}%
}
\end{table*}

\begin{figure}
    \centering
    
    \includegraphics[scale=0.43]{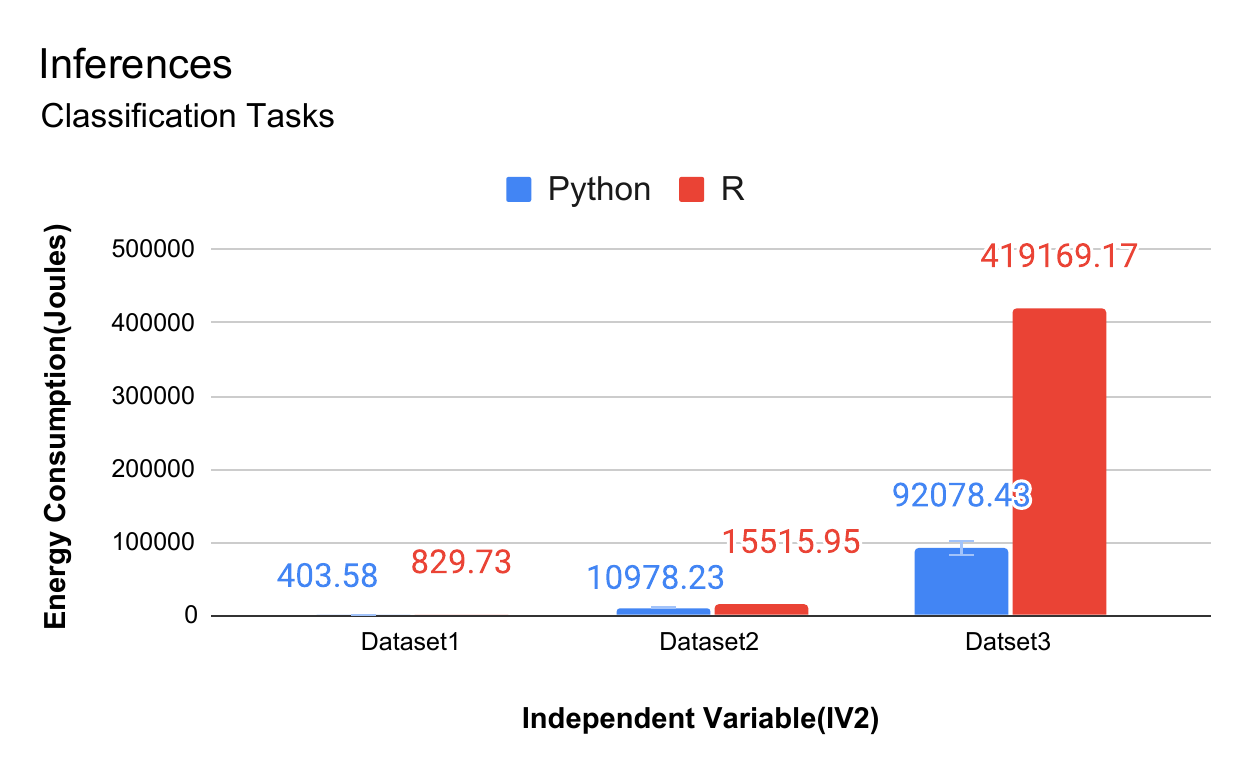}
    \caption{Cumulative Median Energy Consumption of Classification Tasks for Inferences Phase}
    \label{fig:classification2}
\end{figure}
\textbf{C. Findings for  RQ2.1: Comparing Python and R for Energy Efficiency and Run-Time Performance in Machine Learning Model Training for classification Tasks.}

The subsequent research inquiry focuses on discerning the programming language, Python or R, that exhibits superior energy efficiency and run-time performance when training models across various classification tasks. Table \ref{tab:my-table3} illustrates the experimental findings, indicating that, except for Decision Tree classification, Python exhibits significantly higher energy and run-time efficiency compared to R. For instance, in Dataset1, Python demonstrates energy efficiencies of 91\%, 98\%, 83.5\%, and 77.7\% for Gaussian Naive Bayes, Logistic Regression, Random Forest, and Support Vector Machine, respectively, when contrasted with R. 
Similarly, in Dataset2, Python shows better energy efficiency across all algorithms except for Decision Tree (where R is 99.5\% more energy efficient), with energy efficiency gains of 98.5\%, 91.3\%, 86\%, and 54\% for Logistic Regression, Gaussian Naive Bayes, Support Vector Machine, and Random Forest, respectively. Python exhibits \textit{60x} improvement in run-time performance for Logistic Regression, while R outperforms Python by a factor of 11 in Decision Tree classification.

In Dataset3, the trend continues, with Python showing considerable energy efficiency gains for most algorithms. For example, Python exhibits 92.6\% more energy efficiency in Support Vector Machine and is 81.4\% more efficient in Random Forest compared to R. However, R demonstrates a 96.1\% higher energy efficiency in Decision Tree classification for Dataset3. Across the board, Python continues to excel in run-time performance, outperforming R by a substantial margin in most tasks, particularly in Logistic Regression and Support Vector Machine, where Python is 85x and 10x faster than R, respectively.

\textbf{D. Findings for RQ2.2: Comparing Python and R for Energy Efficiency and Run-Time Performance in Machine Learning Model Inferences for classification Tasks.}


The final research question aims to distinguish between Python and R in terms of their energy and run-time efficiency during inference across diverse classification tasks. The observations from the results for classification tasks are available in Table \ref{tab:my-table3}. When employing Dataset1, for the Decision Tree and Gaussian Naive Bayes models, R displays notably lower energy consumption and run-time than Python. Particularly, R showcases 98.7\% higher energy efficiency and \textit{86.5x} run-time performance speed up for the Decision Tree inferences. However, Random Forest, Logistic Regression, and SVM exhibit higher energy consumption and run-time in R than in Python. For instance, when utilizing the Random Forest algorithm, Python shows a 90.8\% reduction in energy consumption.

Conversely, when we shift to Dataset2, Python surpasses R in terms of energy efficiency for Gaussian Naive Bayes inferences. However, interestingly, R maintains better runtime performance. Despite R maintaining its energy efficiency for the Decision Tree at 99.8\%, Python proves to be more energy-efficient for all other algorithms, with variations ranging from 24.3\% (Gaussian Naive Bayes) to 95.3\% (Random Forest). In Dataset3, Python continues to demonstrate significant energy efficiency during inferences, being 73.8\% more efficient than R. Notably, Python is 91.3\% more energy efficient in Logistic Regression, 96.7\% more efficient in Random Forest, and 97.8\% more efficient in Support Vector Machine during the inference phase. However, R outperforms Python by 86.2\% in Decision Tree classification inferences.


\begin{tcolorbox}[colback=blue!10!white, colframe=blue!50!black, title=Key Insights from RQ2.1 and RQ2.2]
\small
\begin{itemize}
\item Python demonstrates higher energy and run-time efficiency over R in both the model training and inference phases, with the exception of Decision Tree classification. 

\item While R shows higher energy efficiency for Gaussian Naive Bayes during inferences using Dataset1, Python provides more energy efficiency for Dataset2.

\item During Logistic Regression model training, Python demonstrates up to 98.5\% higher energy efficiency, while in the inference phase for Random Forest, it achieves up to 95.5\% greater efficiency.

\item In Dataset3, Python shows significant energy savings, being 97.8\% more energy efficient than R in Support Vector Machine classification during the inference phase, while R remains 86.2\% more efficient than Python in Decision Tree classification inferences.
\end{itemize}

\end{tcolorbox}

These findings demonstrate significant differences in the overall energy usage between Python and R across all datasets. For model training, \textbf{Python is observed to be 65.5\%, 63.7\% and 84.24\% more energy efficient} than R for Dataset1, Dataset2 and Dataset3 respectively, as depicted in Fig. \ref{fig:classification1}. 
In terms of inferences, for Dataset1, \textbf{Python consumes less than half (48.6\%) the energy consumed by R}. For Dataset2, \textbf{Python’s energy efficiency is noted at 29.2\%}, as shown in Fig. \ref{fig:classification2}. In Dataset3, Python maintains a significant advantage, being \textbf{78\% more energy efficient}, consuming \textbf{92,078.43 Joules compared to R's 419,169.17 Joules}.

With an increase in the number of data points from Dataset1 to Dataset2, \textbf{Python and R's energy consumption increases by 23.5 times and 22.4 times}, respectively, during model training for classification tasks. The most rapid increase in energy consumption is observed for Support Vector Machine classification, with a \textbf{20-fold increase for Python} and a \textbf{32-fold increase for R}. Moving from Dataset2 to Dataset3, \textbf{Python's energy consumption increases by 8.5 times}, while \textbf{R's energy consumption grows by 19.5 times}, showcasing a significant disparity.

During the inference phase, \textbf{Python's energy consumption increases by 27 times}, while for \textbf{R}, it rises by \textbf{18.7 times} from Dataset1 to Dataset3. Notably, Python's energy consumption tends to escalate more gradually with larger datasets, particularly for classification tasks, while \textbf{R shows a steep and more pronounced increase in both the model training and inference phases as the dataset size grows}.



\textbf{\textit{Data analysis:}}
The p-values from Wilcoxon Signed Rank Test\cite{woolson2007wilcoxon} presented in Table \ref{tab:my-table2} and Table \ref{tab:my-table3} indicate significant differences in energy consumption values of regression tasks, with a large effect size observed in \textbf{28 out of 30 cases} (93.33\%). Similarly, for classification tasks, the results show statistically different energy consumption outcomes in \textbf{29 out of 30 cases} (96.67\%). \textit{Overall, the energy consumption values for both regression and classification tasks, when implemented in R and Python, exhibit \textbf{statistical differences in 95\%} of the cases.} These findings emphasize the critical role of the implementation language in optimizing energy usage in machine learning tasks, highlighting its potential implications for computational efficiency and resource management.

We calculated and reported standard deviation (SD) and coefficient of variation (CV) percentage values for each task, along with mean values, in the replication package. These metrics provide insights into the spread of energy consumption and run-time readings from the mean. A high SD indicates greater variability, while a low SD suggests that readings are closely clustered around the mean.

We analyzed CV percentage values, calculated as the standard deviation-to-mean ratio multiplied by 100. This statistic is useful for comparing the degree of variation from one data series to another, even if the means are drastically different from one another\cite{abdi2010coefficient}. The average CV percentage values are provided in Table \ref{tab:cv}. Notably, the highest CV percentage was found for Python Gaussian Naive Bayes during the inference phase (166.67\%), while for energy consumption, the highest CV percentage is observed for Logistic Regression inferences (PKG-156.875\%). These values indicate that the SD is larger than the mean, signifying significant data spread, which could be attributed to potential noise in measurements as described in Section \ref{threats}. Despite occasional higher variability, most CV values are small, as indicated by the average CV values shown in Table \ref{tab:cv}, ensuring measurement reliability and enabling robust analysis and interpretation of the data.

\begin{table}[]
\small
\centering
\caption{Average Coefficient of Variation (CV) percentage values for the measurement readings}
\label{tab:cv}
\begin{tabular}{|l|l|l|l|}
\hline
 & Run-Time & PKG & DRAM \\ \hline
Regression Tasks & 13.84 & 7.39 & 18.33 \\ \hline
Classification Tasks & 24.42 & 9.62 & 10.03 \\ \hline
\end{tabular}
\end{table}


\textbf{\textit{Energy consumption and run-time correlation: }}Looking closely at Tables \ref{tab:my-table2} and \ref{tab:my-table3}, we observe that runtime performance generally follows a similar pattern as energy efficiency for both regression and classification tasks. To statistically assess the correlation between the two dependent variables, Energy Consumption (DV1) and Run-Time (DV2), we conducted Spearman's Correlation Test \cite{myers2004spearman} and reported the results in Table \ref{tab:my-table-corr}. The findings indicate that, on average, Python exhibits a strong positive correlation(0.70 $\leq$ $\rho$ $\leq$ 0.89), while R demonstrates a very strong positive correlation(0.90 $\leq$ $\rho$ $\leq$ 1) across all tasks. 

\begin{table*}[]
\small
\centering
\caption{Energy Consumption and Run-Time Correlation Test Results}
\vspace{2mm}
\label{tab:my-table-corr}
\resizebox{\textwidth}{!}{%
\begin{tabular}{|c|cccc|cccc|}
\hline
\multirow{3}{*}{}               & \multicolumn{4}{c|}{\textbf{Regression Tasks}}                                                                                                                                                                                                                       & \multicolumn{4}{c|}{\textbf{Classification Tasks}}                                                                                                                                                                                                                   \\ \cline{2-9} 
                                & \multicolumn{2}{c|}{\textbf{Python}}                                                                                                        & \multicolumn{2}{c|}{\textbf{R}}                                                                                        & \multicolumn{2}{c|}{\textbf{Python}}                                                                                                        & \multicolumn{2}{c|}{\textbf{R}}                                                                                        \\ \cline{2-9} 
                                & \multicolumn{1}{c|}{\textbf{\begin{tabular}[c]{@{}c@{}}Correlation \\ Coefficient (\textbf{$\boldsymbol{\rho}$})\end{tabular}}} & \multicolumn{1}{c|}{\textbf{p-value}} & \multicolumn{1}{c|}{\textbf{\begin{tabular}[c]{@{}c@{}}Correlation \\ Coefficient (\textbf{$\boldsymbol{\rho}$})\end{tabular}}} & \textbf{p-value} & \multicolumn{1}{c|}{\textbf{\begin{tabular}[c]{@{}c@{}}Correlation \\ Coefficient (\textbf{$\boldsymbol{\rho}$})\end{tabular}}} & \multicolumn{1}{c|}{\textbf{p-value}} & \multicolumn{1}{c|}{\textbf{\begin{tabular}[c]{@{}c@{}}Correlation \\ Coefficient (\textbf{$\boldsymbol{\rho}$})\end{tabular}}} & \textbf{p-value} \\ \hline
\textit{\textbf{PKG-Run Time}}  & \multicolumn{1}{c|}{0.86}                                                                           & \multicolumn{1}{c|}{1.30e-06}         & \multicolumn{1}{c|}{0.97}                                                                           & 2.97e-12         & \multicolumn{1}{c|}{0.89}                                                                           & \multicolumn{1}{c|}{1.85e-07}         & \multicolumn{1}{c|}{0.93}                                                                           & 2.21e-09         \\ \hline
\textit{\textbf{DRAM-Run Time}} & \multicolumn{1}{c|}{0.89}                                                                           & \multicolumn{1}{c|}{2.02e-07}         & \multicolumn{1}{c|}{0.96}                                                                           & 4.72e-11         & \multicolumn{1}{c|}{0.94}                                                                           & \multicolumn{1}{c|}{3.11e-10}         & \multicolumn{1}{c|}{0.92}                                                                           & 6.58e-09         \\ \hline
\end{tabular}%
}
\end{table*}

\subsection{Comparative Analysis of Python-scratch Implementations and R Implementations}






There are multiple research studies in the literature on comparing energy consumption across various programming languages\cite{pereira2021ranking} and libraries\cite{shanbhag2023exploratory}. For example, Georgiou et al.\cite{georgiou2017analyzing} investigated the energy consumption of 14 programming languages across multiple computational tasks from the Rosetta Code repository implementations, while Verdecchia et al. \cite{verdecchia2022data} analyzed the energy consumption of machine learning algorithms using \textit{scikit-learn} implementations. Shanbagh et al. \cite{shanbhag2023exploratory} and Nahrstedt et al. \cite{nahrstedt2024empirical} have focused on pre-processing tasks across different dataframe libraries. While it is a common practice in the literature for these energy consumption studies to rely on the standard implementations, in this study our goal was to dive deeper and investigate the energy consumption based on scratch implementations with minimal library dependencies. By reducing reliance on pre-optimized libraries, we intend to understand the consistency of the energy comparison results with respect to scratch implementation.

In this study, we utilized \textit{scikit-learn} for Python and packages from the \textit{CRAN} repository for R to perform model training and inferences on various regression and classification tasks. Any machine learning model can be implemented in multiple ways even within the same programming language. To understand if there is consistency amongst various implementations of classification and regression tasks in ML, we replicated the experiments as described in Section \ref{result1} using scratch implementations.

Specifically, we sought Python implementations of regression and classification tasks that adhered to classical machine learning techniques without relying on optimized libraries like \textit{scikit-learn} and have identified  the \textbf{\texttt{ML-From-Scratch}}\footnote{\url{https://github.com/eriklindernoren/ML-From-Scratch}} repository on GitHub, which aligns with our experimental design. This repository has over \textbf{23.9K stars} and 4.6K forks and offers raw implementations of various machine learning algorithms. Unlike the commonly used implementations in Python, R predominantly relies on the CRAN ecosystem which offers implementations via multiple libraries for machine learning algorithms. Hence we limited our preliminary explorations to Python based scratch implementations.

In Table \ref{tab:scratch-regression} and Table \ref{tab:scratch-classification}, we present the comparative mean results of regression and classification tasks using the scratch implementations. Finally, in Table \ref{tab:regression-final} and Table \ref{tab:classification-final} we summarize the key findings, highlighting how the scratch implementations \textbf{align} with or \textbf{diverge} from the \textit{scikit-learn} results for regression and classification tasks, respectively. These results validate the trends observed with the use of \textit{scikit-learn}, with only \textbf{one exception} (Gaussian Naive Bayes Classification during Inferences on Dataset2), as highlighted in Table \ref{tab:classification-final}. In all other cases (\textbf{59 out of 60} cases observed), the scratch implementations follow the same pattern as those observed in using \textit{scikit-learn}. This consistency reinforces the conclusions drawn from the initial experiments using \textit{scikit-learn} and demonstrates that the observed trends are reliable, independent of the library used.

One additional finding from the results of Table \ref{tab:scratch-regression} and Table \ref{tab:scratch-classification} is that, except for the inferences of Regression-Dataset1 involving Gaussian Regression, Linear Regression, and Support Vector Machine, all other instances show that the scratch implementations consumed more energy compared to the scikit-learn implementations. The \textbf{highest percentage increase} is observed in Gaussian Naive Bayes Classification-Dataset3 during model training, with a significant increase of \textbf{934.44\%}. The \textbf{lowest percentage increase} is noted in Gaussian Regression-Dataset2 during model training, with a modest increase of \textbf{4.89\%}. The reason scratch implementations generally consume more energy than scikit-learn could be attributed to several factors: 

\begin{table*}[]
\centering
\caption{Mean Energy Consumption values (Joules) and Run Time (Seconds) for Regression Tasks-Using Scratch Implementations of Python}
\vspace{2mm}
\label{tab:scratch-regression}
\resizebox{\textwidth}{!}{%
\begin{tabular}{lllllllllllllll}
\hline
\multicolumn{1}{|l|}{\multirow{3}{*}{\textbf{Machine Learning Algorithms(IV1)}}} & \multicolumn{7}{l|}{\textbf{Model Training Phase}} & \multicolumn{7}{l|}{\textbf{Inferences Phase}} \\ \cline{2-15} 
\multicolumn{1}{|l|}{} & \multicolumn{3}{l|}{\textbf{Python}} & \multicolumn{3}{l|}{\textbf{R}} & \multicolumn{1}{l|}{\multirow{2}{*}{\textbf{p-value}}} & \multicolumn{3}{l|}{\textbf{Python}} & \multicolumn{3}{l|}{\textbf{R}} & \multicolumn{1}{l|}{\multirow{2}{*}{\textbf{p-value}}} \\ \cline{2-7} \cline{9-14}
\multicolumn{1}{|l|}{} & \multicolumn{1}{l|}{\textbf{Run-Time}} & \multicolumn{1}{l|}{\textbf{Package}} & \multicolumn{1}{l|}{\textbf{DRAM}} & \multicolumn{1}{l|}{\textbf{Run-Time}} & \multicolumn{1}{l|}{\textbf{Package}} & \multicolumn{1}{l|}{\textbf{DRAM}} & \multicolumn{1}{l|}{} & \multicolumn{1}{l|}{\textbf{Run-Time}} & \multicolumn{1}{l|}{\textbf{Package}} & \multicolumn{1}{l|}{\textbf{DRAM}} & \multicolumn{1}{l|}{\textbf{Run-Time}} & \multicolumn{1}{l|}{\textbf{Package}} & \multicolumn{1}{l|}{\textbf{DRAM}} & \multicolumn{1}{l|}{} \\ \hline
\multicolumn{15}{|c|}{\textbf{Dataset1(IV2)}} \\ \hline
\multicolumn{1}{|l|}{Decision Tree} & \multicolumn{1}{l|}{0.3} & \multicolumn{1}{l|}{12} & \multicolumn{1}{l|}{7.5} & \multicolumn{1}{l|}{0.02} & \multicolumn{1}{l|}{2.15} & \multicolumn{1}{l|}{0.22} & \multicolumn{1}{l|}{0.002} & \multicolumn{1}{l|}{0.093} & \multicolumn{1}{l|}{1.31} & \multicolumn{1}{l|}{0.11} & \multicolumn{1}{l|}{0.003} & \multicolumn{1}{l|}{0.37} & \multicolumn{1}{l|}{0.03} & \multicolumn{1}{l|}{0.008} \\ \hline
\multicolumn{1}{|l|}{Gaussian Regression} & \multicolumn{1}{l|}{60.77} & \multicolumn{1}{l|}{9000.77} & \multicolumn{1}{l|}{552.9} & \multicolumn{1}{l|}{0.38} & \multicolumn{1}{l|}{36.92} & \multicolumn{1}{l|}{3.64} & \multicolumn{1}{l|}{0.002} & \multicolumn{1}{l|}{19.7} & \multicolumn{1}{l|}{1263} & \multicolumn{1}{l|}{98.5} & \multicolumn{1}{l|}{0.023} & \multicolumn{1}{l|}{2.39} & \multicolumn{1}{l|}{0.24} & \multicolumn{1}{l|}{0.002} \\ \hline
\multicolumn{1}{|l|}{Linear Regression} & \multicolumn{1}{l|}{0.08} & \multicolumn{1}{l|}{4.12} & \multicolumn{1}{l|}{0.57} & \multicolumn{1}{l|}{0.18} & \multicolumn{1}{l|}{17.59} & \multicolumn{1}{l|}{1.71} & \multicolumn{1}{l|}{0.002} & \multicolumn{1}{l|}{0.011} & \multicolumn{1}{l|}{0.42} & \multicolumn{1}{l|}{0.06} & \multicolumn{1}{l|}{0.18} & \multicolumn{1}{l|}{17.5} & \multicolumn{1}{l|}{1.75} & \multicolumn{1}{l|}{0.002} \\ \hline
\multicolumn{1}{|l|}{Neural Network Regression} & \multicolumn{1}{l|}{25.99} & \multicolumn{1}{l|}{7865.44} & \multicolumn{1}{l|}{311.65} & \multicolumn{1}{l|}{0.58} & \multicolumn{1}{l|}{56.79} & \multicolumn{1}{l|}{5.88} & \multicolumn{1}{l|}{0.002} & \multicolumn{1}{l|}{1.87} & \multicolumn{1}{l|}{4.65} & \multicolumn{1}{l|}{1.28} & \multicolumn{1}{l|}{0.002} & \multicolumn{1}{l|}{0.32} & \multicolumn{1}{l|}{0.03} & \multicolumn{1}{l|}{0.002} \\ \hline
\multicolumn{1}{|l|}{Support Vector Machine} & \multicolumn{1}{l|}{43.21} & \multicolumn{1}{l|}{1278.17} & \multicolumn{1}{l|}{98.2} & \multicolumn{1}{l|}{19.69} & \multicolumn{1}{l|}{1923} & \multicolumn{1}{l|}{175} & \multicolumn{1}{l|}{0.002} & \multicolumn{1}{l|}{5.43} & \multicolumn{1}{l|}{146} & \multicolumn{1}{l|}{14.9} & \multicolumn{1}{l|}{0.003} & \multicolumn{1}{l|}{0.37} & \multicolumn{1}{l|}{0.03} & \multicolumn{1}{l|}{0.002} \\ \hline
\multicolumn{15}{|c|}{\textbf{Dataset2(IV2)}} \\ \hline
\multicolumn{1}{|l|}{Decision Tree} & \multicolumn{1}{l|}{1.13} & \multicolumn{1}{l|}{19.8} & \multicolumn{1}{l|}{10.3} & \multicolumn{1}{l|}{0.143} & \multicolumn{1}{l|}{10.65} & \multicolumn{1}{l|}{1.36} & \multicolumn{1}{l|}{0.002} & \multicolumn{1}{l|}{0.022} & \multicolumn{1}{l|}{2.4} & \multicolumn{1}{l|}{1.5} & \multicolumn{1}{l|}{0.009} & \multicolumn{1}{l|}{0.73} & \multicolumn{1}{l|}{0.11} & \multicolumn{1}{l|}{0.002} \\ \hline
\multicolumn{1}{|l|}{Gaussian Regression} & \multicolumn{1}{l|}{721.9} & \multicolumn{1}{l|}{73254} & \multicolumn{1}{l|}{1035.49} & \multicolumn{1}{l|}{26.398} & \multicolumn{1}{l|}{7124.58} & \multicolumn{1}{l|}{573.21} & \multicolumn{1}{l|}{0.002} & \multicolumn{1}{l|}{215.6} & \multicolumn{1}{l|}{12450} & \multicolumn{1}{l|}{1440} & \multicolumn{1}{l|}{3.291} & \multicolumn{1}{l|}{322.63} & \multicolumn{1}{l|}{31.2} & \multicolumn{1}{l|}{0.002} \\ \hline
\multicolumn{1}{|l|}{Linear Regression} & \multicolumn{1}{l|}{0.022} & \multicolumn{1}{l|}{1134.34} & \multicolumn{1}{l|}{61.9} & \multicolumn{1}{l|}{20.997} & \multicolumn{1}{l|}{38485.03} & \multicolumn{1}{l|}{2533.54} & \multicolumn{1}{l|}{0.002} & \multicolumn{1}{l|}{0.014} & \multicolumn{1}{l|}{57.6} & \multicolumn{1}{l|}{8.4} & \multicolumn{1}{l|}{19.727} & \multicolumn{1}{l|}{501.42} & \multicolumn{1}{l|}{39.7} & \multicolumn{1}{l|}{0.002} \\ \hline
\multicolumn{1}{|l|}{Neural Network Regression} & \multicolumn{1}{l|}{6.1} & \multicolumn{1}{l|}{7643.33} & \multicolumn{1}{l|}{922.4} & \multicolumn{1}{l|}{16.32} & \multicolumn{1}{l|}{356.38} & \multicolumn{1}{l|}{46.34} & \multicolumn{1}{l|}{0.002} & \multicolumn{1}{l|}{0.067} & \multicolumn{1}{l|}{12.8} & \multicolumn{1}{l|}{5.2} & \multicolumn{1}{l|}{0.008} & \multicolumn{1}{l|}{1.45} & \multicolumn{1}{l|}{0.83} & \multicolumn{1}{l|}{0.002} \\ \hline
\multicolumn{1}{|l|}{Support Vector Machine} & \multicolumn{1}{l|}{119.7} & \multicolumn{1}{l|}{5109.87} & \multicolumn{1}{l|}{6348} & \multicolumn{1}{l|}{9.366} & \multicolumn{1}{l|}{55962.95} & \multicolumn{1}{l|}{4545.79} & \multicolumn{1}{l|}{0.002} & \multicolumn{1}{l|}{63.4} & \multicolumn{1}{l|}{4780} & \multicolumn{1}{l|}{785} & \multicolumn{1}{l|}{2.115} & \multicolumn{1}{l|}{28.76} & \multicolumn{1}{l|}{2.73} & \multicolumn{1}{l|}{0.002} \\ \hline
\multicolumn{15}{c}{\textbf{Dataset3(IV2)}} \\ \hline
\multicolumn{1}{|l|}{Decision Tree} & \multicolumn{1}{l|}{5.7} & \multicolumn{1}{l|}{35.9} & \multicolumn{1}{l|}{5.9} & \multicolumn{1}{l|}{2.7} & \multicolumn{1}{l|}{214.3} & \multicolumn{1}{l|}{27.2} & \multicolumn{1}{l|}{0.002} & \multicolumn{1}{l|}{0.42} & \multicolumn{1}{l|}{7.3} & \multicolumn{1}{l|}{3.2} & \multicolumn{1}{l|}{0.32} & \multicolumn{1}{l|}{27.4} & \multicolumn{1}{l|}{3.1} & \multicolumn{1}{l|}{0.002} \\ \hline
\multicolumn{1}{|l|}{Gaussian Regression} & \multicolumn{1}{l|}{891.4} & \multicolumn{1}{l|}{115012.14} & \multicolumn{1}{l|}{8351.39} & \multicolumn{1}{l|}{26.421} & \multicolumn{1}{l|}{61444.22} & \multicolumn{1}{l|}{15756.62} & \multicolumn{1}{l|}{0.002} & \multicolumn{1}{l|}{224.3} & \multicolumn{1}{l|}{210450} & \multicolumn{1}{l|}{26000} & \multicolumn{1}{l|}{122.43} & \multicolumn{1}{l|}{12020.58} & \multicolumn{1}{l|}{1160.65} & \multicolumn{1}{l|}{0.002} \\ \hline
\multicolumn{1}{|l|}{Linear Regression} & \multicolumn{1}{l|}{0.14} & \multicolumn{1}{l|}{6.71} & \multicolumn{1}{l|}{1.3} & \multicolumn{1}{l|}{20.999} & \multicolumn{1}{l|}{78485.03} & \multicolumn{1}{l|}{12533.54} & \multicolumn{1}{l|}{0.002} & \multicolumn{1}{l|}{0.033} & \multicolumn{1}{l|}{1.75} & \multicolumn{1}{l|}{0.28} & \multicolumn{1}{l|}{0.63} & \multicolumn{1}{l|}{541.3} & \multicolumn{1}{l|}{43.5} & \multicolumn{1}{l|}{0.002} \\ \hline
\multicolumn{1}{|l|}{Neural Network Regression} & \multicolumn{1}{l|}{29.5} & \multicolumn{1}{l|}{8159.2} & \multicolumn{1}{l|}{315.7} & \multicolumn{1}{l|}{92.5} & \multicolumn{1}{l|}{27685.9} & \multicolumn{1}{l|}{3421.5} & \multicolumn{1}{l|}{0.002} & \multicolumn{1}{l|}{0.29} & \multicolumn{1}{l|}{13.9} & \multicolumn{1}{l|}{2.1} & \multicolumn{1}{l|}{732.651} & \multicolumn{1}{l|}{7276.54} & \multicolumn{1}{l|}{119.21} & \multicolumn{1}{l|}{0.002} \\ \hline
\multicolumn{1}{|l|}{Support Vector Machine} & \multicolumn{1}{l|}{620.2} & \multicolumn{1}{l|}{53800.23} & \multicolumn{1}{l|}{11120.54} & \multicolumn{1}{l|}{46.8} & \multicolumn{1}{l|}{92420.8} & \multicolumn{1}{l|}{7525.5} & \multicolumn{1}{l|}{0.002} & \multicolumn{1}{l|}{213.6} & \multicolumn{1}{l|}{24560} & \multicolumn{1}{l|}{1900} & \multicolumn{1}{l|}{78.5976} & \multicolumn{1}{l|}{79219.12} & \multicolumn{1}{l|}{348.38} & \multicolumn{1}{l|}{0.002} \\ \hline
\end{tabular}%
}
\end{table*}

\begin{table*}[]
\centering
\caption{Mean Energy Consumption values (Joules) and Run Time (Seconds) for Classification Tasks-Using Scratch Implementations of Python}
\vspace{2mm}
\label{tab:scratch-classification}
\resizebox{\textwidth}{!}{%
\begin{tabular}{|lllllllllllllll|}
\hline
\multicolumn{1}{|l|}{\multirow{3}{*}{\textbf{Machine Learning Algorithms(IV1)}}} & \multicolumn{7}{l|}{\textbf{Model Training Phase}} & \multicolumn{7}{l|}{\textbf{Inferences Phase}} \\ \cline{2-15} 
\multicolumn{1}{|l|}{} & \multicolumn{3}{l|}{\textbf{Python}} & \multicolumn{3}{l|}{\textbf{R}} & \multicolumn{1}{l|}{\multirow{2}{*}{\textbf{p-value}}} & \multicolumn{3}{l|}{Python} & \multicolumn{3}{l|}{R} & \multirow{2}{*}{\textbf{p-value}} \\ \cline{2-7} \cline{9-14}
\multicolumn{1}{|l|}{} & \textbf{Run-Time} & \textbf{Package} & \textbf{DRAM} & \textbf{Run-Time} & \textbf{Package} & \multicolumn{1}{l|}{\textbf{DRAM}} & \multicolumn{1}{l|}{} & \textbf{Run-Time} & \textbf{Package} & \textbf{DRAM} & \textbf{Run-Time} & \textbf{Package} & \multicolumn{1}{l|}{\textbf{DRAM}} &  \\ \hline
\multicolumn{15}{|c|}{\textbf{Dataset1(IV2)}} \\ \hline
\multicolumn{1}{|l|}{Decision Tree} & \multicolumn{1}{l|}{7.8} & \multicolumn{1}{l|}{671.54} & \multicolumn{1}{l|}{65.4} & \multicolumn{1}{l|}{0.476} & \multicolumn{1}{l|}{46.34} & \multicolumn{1}{l|}{4.3} & \multicolumn{1}{l|}{0.002} & \multicolumn{1}{l|}{2.8} & \multicolumn{1}{l|}{230} & \multicolumn{1}{l|}{22.5} & \multicolumn{1}{l|}{0.022} & \multicolumn{1}{l|}{2.25} & \multicolumn{1}{l|}{0.23} & 0.002 \\ \hline
\multicolumn{1}{|l|}{Gaussian Naive Bayes} & \multicolumn{1}{l|}{0.017} & \multicolumn{1}{l|}{1.03} & \multicolumn{1}{l|}{0.14} & \multicolumn{1}{l|}{0.019} & \multicolumn{1}{l|}{1.99} & \multicolumn{1}{l|}{\textbf{0.19}} & \multicolumn{1}{l|}{0.002} & \multicolumn{1}{l|}{0.009} & \multicolumn{1}{l|}{0.49} & \multicolumn{1}{l|}{0.07} & \multicolumn{1}{l|}{0.001} & \multicolumn{1}{l|}{0.21} & \multicolumn{1}{l|}{0.02} & 0.002 \\ \hline
\multicolumn{1}{|l|}{Logistic Regression} & \multicolumn{1}{l|}{0.063} & \multicolumn{1}{l|}{6.32} & \multicolumn{1}{l|}{0.48} & \multicolumn{1}{l|}{2.345} & \multicolumn{1}{l|}{226.63} & \multicolumn{1}{l|}{21.84} & \multicolumn{1}{l|}{0.002} & \multicolumn{1}{l|}{0.015} & \multicolumn{1}{l|}{1.1} & \multicolumn{1}{l|}{0.12} & \multicolumn{1}{l|}{0.03} & \multicolumn{1}{l|}{3.09} & \multicolumn{1}{l|}{0.32} & 0.002 \\ \hline
\multicolumn{1}{|l|}{Random Forest} & \multicolumn{1}{l|}{2.8} & \multicolumn{1}{l|}{212.4} & \multicolumn{1}{l|}{19.3} & \multicolumn{1}{l|}{8.173} & \multicolumn{1}{l|}{794.97} & \multicolumn{1}{l|}{66.18} & \multicolumn{1}{l|}{0.002} & \multicolumn{1}{l|}{0.095} & \multicolumn{1}{l|}{5.5} & \multicolumn{1}{l|}{0.7} & \multicolumn{1}{l|}{0.406} & \multicolumn{1}{l|}{39.15} & \multicolumn{1}{l|}{3.33} & 0.002 \\ \hline
\multicolumn{1}{|l|}{Support Vector Machine} & \multicolumn{1}{l|}{7.9} & \multicolumn{1}{l|}{690.11} & \multicolumn{1}{l|}{67.8} & \multicolumn{1}{l|}{23.48} & \multicolumn{1}{l|}{2343.45} & \multicolumn{1}{l|}{182.67} & \multicolumn{1}{l|}{0.002} & \multicolumn{1}{l|}{2.9} & \multicolumn{1}{l|}{237.2} & \multicolumn{1}{l|}{23.4} & \multicolumn{1}{l|}{7.242} & \multicolumn{1}{l|}{725.12} & \multicolumn{1}{l|}{56.01} & 0.002 \\ \hline
\multicolumn{15}{|c|}{\textbf{Dataset2(IV2)}} \\ \hline
\multicolumn{1}{|l|}{Decision Tree} & \multicolumn{1}{l|}{165.8} & \multicolumn{1}{l|}{14643.31} & \multicolumn{1}{l|}{8200} & \multicolumn{1}{l|}{0.596} & \multicolumn{1}{l|}{57.59} & \multicolumn{1}{l|}{34.1} & \multicolumn{1}{l|}{0.002} & \multicolumn{1}{l|}{57.5} & \multicolumn{1}{l|}{5543.22} & \multicolumn{1}{l|}{3250} & \multicolumn{1}{l|}{0.064} & \multicolumn{1}{l|}{6.5} & \multicolumn{1}{l|}{3.72} & 0.002 \\ \hline
\multicolumn{1}{|l|}{Gaussian Naive Bayes} & \multicolumn{1}{l|}{0.024} & \multicolumn{1}{l|}{1.4} & \multicolumn{1}{l|}{0.68} & \multicolumn{1}{l|}{0.097} & \multicolumn{1}{l|}{9.6} & \multicolumn{1}{l|}{5.66} & \multicolumn{1}{l|}{0.002} & \multicolumn{1}{l|}{0.012} & \multicolumn{1}{l|}{0.65} & \multicolumn{1}{l|}{0.33} & \multicolumn{1}{l|}{0.004} & \multicolumn{1}{l|}{0.49} & \multicolumn{1}{l|}{0.29} & 0.046 \\ \hline
\multicolumn{1}{|l|}{Logistic Regression} & \multicolumn{1}{l|}{0.095} & \multicolumn{1}{l|}{9.2} & \multicolumn{1}{l|}{6.3} & \multicolumn{1}{l|}{24.34} & \multicolumn{1}{l|}{528.34} & \multicolumn{1}{l|}{237} & \multicolumn{1}{l|}{0.002} & \multicolumn{1}{l|}{0.017} & \multicolumn{1}{l|}{0.24} & \multicolumn{1}{l|}{0.31} & \multicolumn{1}{l|}{0.086} & \multicolumn{1}{l|}{5.02} & \multicolumn{1}{l|}{1.3} & 0.002 \\ \hline
\multicolumn{1}{|l|}{Random Forest} & \multicolumn{1}{l|}{9.9} & \multicolumn{1}{l|}{338.12} & \multicolumn{1}{l|}{390} & \multicolumn{1}{l|}{67.23} & \multicolumn{1}{l|}{1037.92} & \multicolumn{1}{l|}{169.45} & \multicolumn{1}{l|}{0.002} & \multicolumn{1}{l|}{0.23} & \multicolumn{1}{l|}{6.3} & \multicolumn{1}{l|}{7.1} & \multicolumn{1}{l|}{3.78} & \multicolumn{1}{l|}{191.65} & \multicolumn{1}{l|}{22.67} & 0.002 \\ \hline
\multicolumn{1}{|l|}{Support Vector Machine} & \multicolumn{1}{l|}{145.3} & \multicolumn{1}{l|}{6112.44} & \multicolumn{1}{l|}{7800} & \multicolumn{1}{l|}{4.041} & \multicolumn{1}{l|}{76371.35} & \multicolumn{1}{l|}{4120.02} & \multicolumn{1}{l|}{0.002} & \multicolumn{1}{l|}{58.9} & \multicolumn{1}{l|}{5158.3} & \multicolumn{1}{l|}{3865.23} & \multicolumn{1}{l|}{4.04} & \multicolumn{1}{l|}{14151.96} & \multicolumn{1}{l|}{1132.35} & 0.002 \\ \hline
\multicolumn{15}{|c|}{\textbf{Dataset3(IV2)}} \\ \hline
\multicolumn{1}{|l|}{Decision Tree} & \multicolumn{1}{l|}{3125.5} & \multicolumn{1}{l|}{145133.1} & \multicolumn{1}{l|}{13000} & \multicolumn{1}{l|}{9.53} & \multicolumn{1}{l|}{921.47} & \multicolumn{1}{l|}{86.31} & \multicolumn{1}{l|}{0.002} & \multicolumn{1}{l|}{1240.9} & \multicolumn{1}{l|}{57237.77} & \multicolumn{1}{l|}{5134.66} & \multicolumn{1}{l|}{1.028} & \multicolumn{1}{l|}{101.53} & \multicolumn{1}{l|}{10.18} & 0.002 \\ \hline
\multicolumn{1}{|l|}{Gaussian Naive Bayes} & \multicolumn{1}{l|}{1.75} & \multicolumn{1}{l|}{92.3} & \multicolumn{1}{l|}{7.1} & \multicolumn{1}{l|}{1.548} & \multicolumn{1}{l|}{153.56} & \multicolumn{1}{l|}{14.69} & \multicolumn{1}{l|}{0.002} & \multicolumn{1}{l|}{0.23} & \multicolumn{1}{l|}{5.1} & \multicolumn{1}{l|}{0.92} & \multicolumn{1}{l|}{0.064} & \multicolumn{1}{l|}{7.8} & \multicolumn{1}{l|}{0.71} & 0.002 \\ \hline
\multicolumn{1}{|l|}{Logistic Regression} & \multicolumn{1}{l|}{0.47} & \multicolumn{1}{l|}{11.9} & \multicolumn{1}{l|}{1.75} & \multicolumn{1}{l|}{105.78} & \multicolumn{1}{l|}{45267.23} & \multicolumn{1}{l|}{1345.78} & \multicolumn{1}{l|}{0.002} & \multicolumn{1}{l|}{0.19} & \multicolumn{1}{l|}{4.7} & \multicolumn{1}{l|}{0.75} & \multicolumn{1}{l|}{15.89} & \multicolumn{1}{l|}{56.55} & \multicolumn{1}{l|}{267.21} & 0.002 \\ \hline
\multicolumn{1}{|l|}{Random Forest} & \multicolumn{1}{l|}{148.7} & \multicolumn{1}{l|}{6850} & \multicolumn{1}{l|}{820} & \multicolumn{1}{l|}{195.43} & \multicolumn{1}{l|}{42356} & \multicolumn{1}{l|}{1876.11} & \multicolumn{1}{l|}{0.002} & \multicolumn{1}{l|}{4.35} & \multicolumn{1}{l|}{162.73} & \multicolumn{1}{l|}{20.42} & \multicolumn{1}{l|}{121.72} & \multicolumn{1}{l|}{1099.34} & \multicolumn{1}{l|}{654.12} & 0.002 \\ \hline
\multicolumn{1}{|l|}{Support Vector Machine} & \multicolumn{1}{l|}{2500.87} & \multicolumn{1}{l|}{246123.25} & \multicolumn{1}{l|}{11500.41} & \multicolumn{1}{l|}{233.582} & \multicolumn{1}{l|}{1396581.56} & \multicolumn{1}{l|}{114462.55} & \multicolumn{1}{l|}{0.002} & \multicolumn{1}{l|}{1280.1} & \multicolumn{1}{l|}{59002.56} & \multicolumn{1}{l|}{4945.33} & \multicolumn{1}{l|}{64.655} & \multicolumn{1}{l|}{386431.3} & \multicolumn{1}{l|}{30540.43} & 0.002 \\ \hline
\end{tabular}%
}
\end{table*}

\begin{table*}[]
\centering
\caption{Energy Efficiency of Python (using \textit{scikit-learn} and scratch implementations) and R During Model Training and Inference Phases for Regression Tasks.}
\vspace{2mm}
\label{tab:regression-final}
\resizebox{\textwidth}{!}{%
\begin{tabular}{|lllll|}
\hline
\multicolumn{1}{|l|}{\textbf{}} & \multicolumn{2}{l|}{\textbf{During Model Training}} & \multicolumn{2}{l|}{\textbf{During Inferences}} \\ \hline
\multicolumn{1}{|l|}{\textbf{ML Algorithmss(IV1)}} & \multicolumn{1}{l|}{\textbf{using scikit-learn for Python}} & \multicolumn{1}{l|}{\textbf{\begin{tabular}[c]{@{}l@{}}using scratch implementations \\ for Python\end{tabular}}} & \multicolumn{1}{l|}{\textbf{\begin{tabular}[c]{@{}l@{}}using scikit-learn for \\ Python\end{tabular}}} & \textbf{\begin{tabular}[c]{@{}l@{}}using scratch implementations \\ for Python\end{tabular}} \\ \hline
\multicolumn{5}{|c|}{\textbf{Dataset1 (IV2)}} \\ \hline
\multicolumn{1}{|l|}{\textbf{Decision Tree}} & \multicolumn{1}{l|}{R} & \multicolumn{1}{l|}{R} & \multicolumn{1}{l|}{R} & R \\ \hline
\multicolumn{1}{|l|}{\textbf{Gaussian Regression}} & \multicolumn{1}{l|}{R} & \multicolumn{1}{l|}{R} & \multicolumn{1}{l|}{R} & R \\ \hline
\multicolumn{1}{|l|}{\textbf{Linear Regression}} & \multicolumn{1}{l|}{Python} & \multicolumn{1}{l|}{Python} & \multicolumn{1}{l|}{Python} & Python \\ \hline
\multicolumn{1}{|l|}{\textbf{Neural Network Regression}} & \multicolumn{1}{l|}{R} & \multicolumn{1}{l|}{R} & \multicolumn{1}{l|}{R} & R \\ \hline
\multicolumn{1}{|l|}{\textbf{Support Vector Machine}} & \multicolumn{1}{l|}{Python} & \multicolumn{1}{l|}{R} & \multicolumn{1}{l|}{Python} & R \\ \hline
\multicolumn{5}{|c|}{\textbf{Dataset2 (IV2)}} \\ \hline
\multicolumn{1}{|l|}{\textbf{Decision Tree}} & \multicolumn{1}{l|}{R} & \multicolumn{1}{l|}{R} & \multicolumn{1}{l|}{R} & R \\ \hline
\multicolumn{1}{|l|}{\textbf{Gaussian Regression}} & \multicolumn{1}{l|}{R} & \multicolumn{1}{l|}{R} & \multicolumn{1}{l|}{R} & R \\ \hline
\multicolumn{1}{|l|}{\textbf{Linear Regression}} & \multicolumn{1}{l|}{Python} & \multicolumn{1}{l|}{Python} & \multicolumn{1}{l|}{Python} & Python \\ \hline
\multicolumn{1}{|l|}{\textbf{Neural Network Regression}} & \multicolumn{1}{l|}{R} & \multicolumn{1}{l|}{R} & \multicolumn{1}{l|}{R} & R \\ \hline
\multicolumn{1}{|l|}{\textbf{Support Vector Machine}} & \multicolumn{1}{l|}{Python} & \multicolumn{1}{l|}{R} & \multicolumn{1}{l|}{Python} & R \\ \hline
\multicolumn{5}{|c|}{\textbf{Dataset3 (IV2)}} \\ \hline
\multicolumn{1}{|l|}{\textbf{Decision Tree}} & \multicolumn{1}{l|}{Python} & \multicolumn{1}{l|}{Python} & \multicolumn{1}{l|}{Python} & Python \\ \hline
\multicolumn{1}{|l|}{\textbf{Gaussian Regression}} & \multicolumn{1}{l|}{R} & \multicolumn{1}{l|}{R} & \multicolumn{1}{l|}{R} & R \\ \hline
\multicolumn{1}{|l|}{\textbf{Linear Regression}} & \multicolumn{1}{l|}{Python} & \multicolumn{1}{l|}{Python} & \multicolumn{1}{l|}{Python} & Python \\ \hline
\multicolumn{1}{|l|}{\textbf{Neural Network Regression}} & \multicolumn{1}{l|}{Python} & \multicolumn{1}{l|}{Python} & \multicolumn{1}{l|}{Python} & Python \\ \hline
\multicolumn{1}{|l|}{\textbf{Support Vector Machine}} & \multicolumn{1}{l|}{Python} & \multicolumn{1}{l|}{Python} & \multicolumn{1}{l|}{Python} & Python \\ \hline
\end{tabular}%
}
\end{table*}

\begin{table*}[]
\centering
\caption{Energy Efficiency of Python (using \textit{scikit-learn} and scratch implementations) and R During Model Training and Inference Phases for Classification Tasks.}
\vspace{2mm}
\label{tab:classification-final}
\resizebox{\textwidth}{!}{%
\begin{tabular}{|lllll|}
\hline
\multicolumn{1}{|l|}{} & \multicolumn{2}{l|}{\textbf{During Model Training}} & \multicolumn{2}{l|}{\textbf{During Inferences}} \\ \cline{2-5} 
\multicolumn{1}{|l|}{\multirow{-2}{*}{\textbf{ML Algorithms (IV1)}}} & \multicolumn{1}{l|}{\textbf{using scikit-learn for Python}} & \multicolumn{1}{l|}{\textbf{\begin{tabular}[c]{@{}l@{}}using scratch implementations\\  for Python\end{tabular}}} & \multicolumn{1}{l|}{\textbf{using scikit-learn for Python}} & \textbf{\begin{tabular}[c]{@{}l@{}}using scratch implementations \\ for Python\end{tabular}} \\ \hline
\multicolumn{5}{|c|}{\textbf{Dataset1 (IV2)}} \\ \hline
\multicolumn{1}{|l|}{\textbf{Decision Tree}} & \multicolumn{1}{l|}{R} & \multicolumn{1}{l|}{R} & \multicolumn{1}{l|}{R} & R \\ \hline
\multicolumn{1}{|l|}{\textbf{Gaussian Naive Bayes}} & \multicolumn{1}{l|}{Python} & \multicolumn{1}{l|}{Python} & \multicolumn{1}{l|}{R} & R \\ \hline
\multicolumn{1}{|l|}{\textbf{Logistic Regression}} & \multicolumn{1}{l|}{Python} & \multicolumn{1}{l|}{Python} & \multicolumn{1}{l|}{Python} & Python \\ \hline
\multicolumn{1}{|l|}{\textbf{Random Forest}} & \multicolumn{1}{l|}{Python} & \multicolumn{1}{l|}{Python} & \multicolumn{1}{l|}{Python} & Python \\ \hline
\multicolumn{1}{|l|}{\textbf{Support Vector Machine}} & \multicolumn{1}{l|}{Python} & \multicolumn{1}{l|}{Python} & \multicolumn{1}{l|}{Python} & Python \\ \hline
\multicolumn{5}{|c|}{\textbf{Dataset2 (IV2)}} \\ \hline
\multicolumn{1}{|l|}{\textbf{Decision Tree}} & \multicolumn{1}{l|}{R} & \multicolumn{1}{l|}{R} & \multicolumn{1}{l|}{R} & R \\ \hline
\multicolumn{1}{|l|}{\textbf{Gaussian Naive Bayes}} & \multicolumn{1}{l|}{Python} & \multicolumn{1}{l|}{Python} & \multicolumn{1}{l|}{\cellcolor[HTML]{C0C0C0}Python} & \cellcolor[HTML]{C0C0C0}R \\ \hline
\multicolumn{1}{|l|}{\textbf{Logistic Regression}} & \multicolumn{1}{l|}{Python} & \multicolumn{1}{l|}{Python} & \multicolumn{1}{l|}{Python} & Python \\ \hline
\multicolumn{1}{|l|}{\textbf{Random Forest}} & \multicolumn{1}{l|}{Python} & \multicolumn{1}{l|}{Python} & \multicolumn{1}{l|}{Python} & Python \\ \hline
\multicolumn{1}{|l|}{\textbf{Support Vector Machine}} & \multicolumn{1}{l|}{Python} & \multicolumn{1}{l|}{Python} & \multicolumn{1}{l|}{Python} & Python \\ \hline
\multicolumn{5}{|c|}{\textbf{Dataset3 (IV2)}} \\ \hline
\multicolumn{1}{|l|}{\textbf{Decision Tree}} & \multicolumn{1}{l|}{R} & \multicolumn{1}{l|}{R} & \multicolumn{1}{l|}{R} & R \\ \hline
\multicolumn{1}{|l|}{\textbf{Gaussian Naive Bayes}} & \multicolumn{1}{l|}{Python} & \multicolumn{1}{l|}{Python} & \multicolumn{1}{l|}{Python} & Python \\ \hline
\multicolumn{1}{|l|}{\textbf{Logistic Regression}} & \multicolumn{1}{l|}{Python} & \multicolumn{1}{l|}{Python} & \multicolumn{1}{l|}{Python} & Python \\ \hline
\multicolumn{1}{|l|}{\textbf{Random Forest}} & \multicolumn{1}{l|}{Python} & \multicolumn{1}{l|}{Python} & \multicolumn{1}{l|}{Python} & Python \\ \hline
\multicolumn{1}{|l|}{\textbf{Support Vector Machine}} & \multicolumn{1}{l|}{Python} & \multicolumn{1}{l|}{Python} & \multicolumn{1}{l|}{Python} & Python \\ \hline
\end{tabular}%
}
\end{table*}

\begin{itemize}

\item \textbf{Optimizations in \textit{scikit-learn}}: 
\textit{scikit-learn} uses Cython and compiled C/C++ code, enabling efficient memory management and faster execution, which reduces energy consumption. Scratch implementations, written in pure Python, are generally slower and less optimized. Similar to how compiler optimizations have been utilized to reduce runtime overhead in performance-critical systems \cite{muller2023capabilities}, scikit-learn achieves significant performance improvements through its use of compiled code, streamlining execution and enhancing efficiency.





\item \textbf{Algorithmic optimizations}: 
\textit{scikit-learn} incorporates optimizations like early stopping, thereby reducing computational steps \cite{ji2023improving}, efficient data handling, and the use of optimized numerical libraries (e.g., BLAS, LAPACK), which are typically absent in scratch implementations, leading to higher energy consumption.
\end{itemize}

\vspace{-2mm}
However, further research can investigate the specific optimizations within \textit{scikit-learn} for each machine learning algorithm to identify areas of energy efficiency and energy-intensive components, providing future direction for improving energy performance


\vspace{-4mm}
\section{Threats to Validity}
\label{threats}

In this section, we delve into potential sources of errors that could pose threats to the results of the empirical study.
\subsection{Internal Validity}

Measuring energy consumption in software systems is challenging and susceptible to threats, particularly from factors like noise, voltage spikes, daemons, and background processes. Moreover, Intel RAPL measures energy consumption at a high sampling rate\cite{desrochers2016validation}. While high sampling rates provide better temporal resolution\cite{hackenberg2013power}, they can also increase the amount of noise in the measurements\cite{ilsche2015power}. Each sample necessitates resources for collection and processing. If training or inference occurs too rapidly, it can introduce extra computational overhead, potentially impacting measurement accuracy.

To mitigate the impact of these issues, we employed several strategies. Firstly, we repeated the same task 10 times and calculated averages from the recorded values and considered the mean values for evaluation. Additionally, we introduced random uniform shuffling while conducting experiments to further reduce the influence of any unnoticed background processes. To counter the potential impact of confounding factors stemming from hardware temperature, we conducted a preliminary CPU-intensive warm-up operation, implementing a Fibonacci sequence calculation for approximately 5 seconds, following the approach outlined by Verdecchia et al. \cite{verdecchia2022data}. This measure guarantees that the hardware does not undergo a ``cold boot''\footnote{\url{https://www.lexicontech.com/resources/blog/cold-boot-vs-warm-boot/}} during the initial execution of our experiment. Moreover, Certain components of a computer system continue to draw higher power even after completing a process, known as power tail states\cite{bornholt2012model}. To address this, we ensured a 30-second idle period before commencing the next task to mitigate the impact of tail states.

Despite our best efforts to minimize these factors' effects, completely eliminating them is challenging, as it would require control over the background operations necessary for running the operating system. To further account for daemons and background threads, we introduced the concept of effective energy consumption ($E_{effective}$), as described in Section \ref{experiments}. While this approach may not be the best method for addressing overheads during energy measurements in a software system, it effectively serves to minimize a portion of the error percentage in the energy readings. Considering hardware limitations, we conducted the experiments on a single system with a standard configuration. It is important to acknowledge the potential for obtaining significantly distinct values on a system with a different configuration.
\vspace{-3mm}
\subsection{External Validity}
An external factor that can potentially influence the outcomes of our study is the choice of datasets. The characteristics of attributes, data types, and dataset sizes can impact the computational workload during task executions, particularly depending on the specific implementations of various machine learning algorithms. Consequently, this can introduce variability in energy measurements. To address this, we conducted our experiments using three datasets with differing sizes and attributes. Our experiments were carried out utilizing established and dependable libraries, namely, \textit{scikit-learn} for Python and the \textit{CRAN repository} for R. Different libraries with distinct implementations may yield different results for various machine learning algorithms and tasks.  Furthermore, both the Python and R languages benefit from active global communities, ensuring continuous optimization and enhancement of the open-source libraries. It is essential to acknowledge that conducting the study with different versions of the same libraries could lead to variations in results. To account for this, we have provided comprehensive information about the versions of the libraries used in this study, as presented in Table \ref{tab:my-table1}. This facilitates the replication of our findings and allows for comparisons with future studies. In addition to this, we have also utilized scratch-level implementations of Python to minimize reliance on library-specific optimizations.



\subsection{Construct Validity}




We have not considered accuracy metrics while comparing the results, since the RQs of the study focuses on the energy cost associated with ML tasks implemented in Python and R. The trade-off is minimal and the scores are comparable in accuracy metrics for most of the algorithms between the same ML model trained using Python and R implementations under same settings\cite{gupta2021performance}. 



 Hyper-parameter Tuning is the process through which user-defined inputs are adjusted in machine learning algorithms to strike a balance between accuracy and generalizability \cite{maher2019smartml}. In this study, we deliberately refrained from pursuing model optimization through hyper-parameter tuning. This decision was made to ensure that our research remained aligned with its core research questions without introducing deviations. To ensure uniformity and minimize external factors' impact, we maintained consistent hyper-parameter values for all algorithms in both classification and regression tasks. The use of tools such as pyJoules and RJoules for energy measurement could have influenced our outcomes. Nonetheless, these tools act as intermediary interfaces retrieving energy measurements from the reliable Intel RAPL \cite{hahnel2012measuring, couto2017towards}.

 \subsection{Conclusion Validity}


A potential threat to the validity of our conclusions may arise from low statistical power. However, we addressed this concern by systematically gathering and analyzing a substantial dataset consisting of 4.8K data points. By employing rigorous statistical tests and methods, we strengthened the validity of our findings, thereby minimizing the risk of drawing erroneous conclusions due to insufficient data.
The focus of our study on examining the energy and run-time efficiency of machine learning tasks using both R and Python's standard libraries. However, it is essential to acknowledge the presence of confounding factors, such as differences in library implementations and algorithmic complexities of ML tasks in \textit{scikit-learn} and \textit{CRAN}, which may introduce bias into our analysis. To minimize this, we also incorporated scratch-level implementations of Python from a reputable GitHub repository, reducing the influence of library-specific optimizations. 

Despite these potential limitations, our rigorous approach to controlled environment experiments, data collection and analysis, coupled with the use of appropriate statistical techniques, strengthens the validity of our conclusions regarding the energy and run-time efficiency of machine learning tasks. 



\section{Discussion}
\label{discussion}
The aim of this study is to comprehend the comparative energy efficiency and run-time performance between the Python and R programming languages across various machine learning tasks, specifically during model training and inference phases.
Our findings indicate that R is more energy- and runtime-efficient in 16 out of 30 regression cases (Table \ref{tab:my-table2}), while Python demonstrates higher energy and runtime efficiency in 23 out of 30 classification cases (Table \ref{tab:my-table3}).



These observed differences might be attributed to certain features intrinsic to specific programming languages, which appear to offer distinct advantages. For example, our findings reveal that model training for SVM regression exhibits superior energy efficiency when executed in R compared to Python. Upon analyzing the source code documentation of the utilized libraries, detailed in Table \ref{tab:my-table1}, we discovered that the \textit{svm()} function within the \textit{e1071} library in R inherently scales the data by default. In contrast, in Python's \textit{scikit-learn}, this scaling process necessitates the use of \textit{fit\_transform}, introducing an additional energy cost. Interestingly, when it comes to inferences, Python demonstrates greater energy efficiency in the same task. This seemingly paradoxical outcome could be explained by the utilization of a 20\% split of the entire dataset for inferences, where the requirement for extensive data scaling is diminished. 

In Section \ref{results}, we reported that for Neural Network Regression, R outperforms Python by 99.16\% in terms of energy efficiency and is 20 times faster. To understand such high variability in performance, we conducted an analysis of the source code documentations of \textit{neuralnet} from CRAN and \textit{neural\_network.MLPRegressor} from \textit{scikit-learn}. Based on our analysis, the observed difference could be attributed to variations in default hyperparameters and optimization settings between \textit{neuralnet} and \textit{MLPRegressor}. However, as discussed in Section \ref{threats}, we opted not to perform hyper\-parameter tuning in this study. Instead, we maintained constant hyper\-parameters across all algorithms.

The results indicate that during model training phase of Logistic Regression, Python offers up to 98.5\% greater energy efficiency and is 60 times faster. Upon analysis of implementations, we could observed that \textit{sklearn.linear\_model.LogisticRegression }in Python utilizes various optimization techniques including gradient descent, stochastic gradient descent, and specialized solvers like liblinear or lbfgs, which are highly optimized and efficient. On the other hand, the \textit{glmnet} package in R implements regularized logistic regression using coordinate descent algorithms with optional cyclical coordinate descent (CD) or sequential strong rule (SSR) algorithms. While these algorithms are effective, they may not be as optimized as those used in \textit{scikit-learn}.

Another observation from our results is that R demonstrates higher energy efficiency for Gaussian Naive Bayes classification during inferences using Dataset1, while Python shows better energy efficient for Dataset2. This result might may appear counter-intuitive  but could stem from the design aspect of Python's \textit{naive\_bayes} package, which generates n-dimensional arrays, aligning well with Python's proficiency in handling matrix computations and optimization tasks, particularly with large datasets. Additionally, \textit{scikit-learn} in Python heavily relies on \textit{NumPy\footnote{\url{https://numpy.org}}} for its operations, which in turn utilizes Cython, a C-based implementation of Python, for certain tasks to enhance performance. Conversely, in R, ML-related libraries often depend on its native or custom data structures. Further investigation into these factors could provide deeper insights into such behaviors.

\textbf{\textit{Potential implications of our findings for practitioners and researchers are as follows:}}
\begin{itemize}
    \item Neither Python nor R stands out as the most energy-efficient programming language across all tasks. The choice should depend on specific task requirements; for instance, Python might not excel in energy efficiency for Decision Tree Regression or Classification due to differences in library implementations (e.g., \textit{rpart} from \textit{CRAN} incorporates pruning \cite{yang2017designing}, while \textit{scikit-learn} does not). However, our experiments have shown that for most of the \textbf{regression} cases, \textbf{R} demonstrates \textbf{better} energy efficiency, while for \textbf{classification tasks}, \textbf{Python} tends to perform more \textbf{efficiently} in terms of both energy consumption and run-time.
     \item The results have demonstrated that the choice of programming language can significantly influence energy consumption by up to \textbf{99.8\%} (e.g., in Gaussian Regression and SVM Regression during model training in Dataset1). 
     \item As the data scales up, for Dataset3 with 1.4M datapoints, a language can show up to \textbf{84.24\%} difference in cumulative energy consumption (Python for classification model training over R on Dataset3). For inferences, the energy consumption difference can be as high as \textbf{98.78\%} (R over Python for regression inferenceson Dataset1). 
    \item The study emphasizes the necessity for enhancing energy efficiency in certain tasks through improvements in the libraries and packages utilized in R and Python. Further research is warranted to delve into the underlying causes and contributors to energy inefficiencies in libraries tailored for specific tasks. Identifying potential workarounds to address these inefficiencies is also essential.
\end{itemize}



\section{Conclusion \& Future work}
\label{conclusion}

In this paper, we have presented an empirical analysis of the energy consumption and run-time performance of machine learning (ML) algorithms implemented in Python and R, focusing on both model training and inference phases. Our aim is to contribute to the understanding of the energy efficiency of programming languages in the context of various stages of the ML process. For the analysis, we evaluated the energy consumption and run-time of 5 regression and 5 classification tasks, detailed in Table \ref{tab:my-table2} and Table \ref{tab:my-table3}, respectively, utilizing standard libraries in Python and R. To measure energy consumption, we employed pyJoules and RJoules, wrappers built on Intel's RAPL interface, following the experimental setup outlined in Section \ref{experiments}. Three datasets from the UCI repository and Kaggle, namely the Adult Dataset, Drug Review dataset and New York City Taxi Trip Duration, were utilized for experimentation, as given in Table \ref{tab:datasets-info} with results summarized in Section \ref{results}. Through addressing RQs about energy and run-time efficiency, we also explored the potential impact of dataset size on energy consumption, and statistically analyzed the correlation between energy consumption and run-time. 
To minimize the effect of library-specific implementations of \textit{scikit-learn} we also conducted experiments using scratch-level implementations for the same regression and classification tasks. This ensures that the trends observed are not influenced solely by the underlying optimizations of any single library. Our examination of the source codes of various algorithms allowed us to gain insights into the performance differences among the algorithms used in the study. We have included some of these findings in Section \ref{discussion}. The findings of this empirical study highlight the significant influence of programming language choice on energy consumption during both the model training and inference phases.

Our objective is to highlight within the research community the significance of studying the energy efficiency of different programming languages within the machine learning (ML) pipeline, particularly during the model training and inference stages. This study serves as a preliminary exploration, with plans to enhance our investigation in several key areas. The current selection of ML tasks is limited; hence, we intend to expand our scope by examining the energy consumption of a broader set of ML tasks across diverse programming languages and system configurations. We also aim to replicate our experiments using larger datasets and multiple systems to better understand emerging trends in energy consumption across various ML operations.

Our study currently focuses on CPU and RAM energy usage, but future work will incorporate GPU-level energy consumption for a more comprehensive analysis. With this study, we aim to provide actionable insights to ML practitioners, enabling informed decisions about energy efficiency when choosing programming languages for different phases of Machine Learning operations (MLOps). As the global ML landscape shifts toward Large Language Models (LLMs) with millions or billions of parameters, there is a growing need for the research community to scrutinize the energy implications of all stages, including pre-training, transfer learning, and fine-tuning. Future work could also explore energy trade-offs involved in hyper-parameter optimization and evaluate the energy and cost implications of decentralized approaches like Federated Learning, thereby contributing significant value to sustainable ML practices.

\section{Data-Availability Statement}
The datasets and scripts used in this study, along with instructions for replication, detailed results, statistical data, and all necessary materials, are provided in the replication package available at \url{https://anonymous.4open.science/r/RvsPython-CFE9/}.

\bibliographystyle{elsarticle-num}
\balance
\bibliography{main}

\end{document}